\documentclass[aps,prl,twocolumn,showpacs,reprint,superscriptaddress,longbibliography,floatfix]{revtex4-1}

\usepackage{amsmath, amsfonts, amssymb, bm}
\usepackage{mathtools}
\usepackage[inline]{enumitem}
\usepackage{graphicx}
\usepackage{xcolor}
\usepackage[breaklinks,hypertexnames=false]{hyperref}
\hypersetup{colorlinks,linkcolor={magenta},citecolor={blue},urlcolor={blue}} 

\usepackage[capitalize]{cleveref}

\usepackage{glossaries}
\glsdisablehyper     
\setacronymstyle{long-short}     

\newacronym{qshi}{QSHI}{quantum spin Hall insulator}  
\newacronym{car}{CAR}{crossed Andreev reflection}
\newacronym{ec}{EC}{electron cotunneling}

\newcommand{\ie}{\textit{i.e., }}

\newcommand{\cf}{\textit{cf. }}

\newcommand{\up}{\uparrow}
\newcommand{\dw}{\downarrow}

\newcommand{\ee}{\mathrm{e}}
\newcommand{\ii}{\mathrm{i}}

\newcommand{\si}{\hat{\sigma}_{0}}
\newcommand{\sx}{\hat{\sigma}_{1}}
\newcommand{\sy}{\hat{\sigma}_{2}}
\newcommand{\sz}{\hat{\sigma}_{3}}

\newcommand{\ti}{\hat{\tau}_{0}}
\newcommand{\tx}{\hat{\tau}_{1}}
\newcommand{\ty}{\hat{\tau}_{2}}
\newcommand{\tz}{\hat{\tau}_{3}}

\DeclarePairedDelimiter\mean{\langle}{\rangle}
\DeclarePairedDelimiter\abs{\lvert}{\rvert}
\DeclareMathOperator{\sgn}{sgn}
\DeclareMathOperator{\real}{Re}
\DeclareMathOperator{\imag}{Im}
\DeclareMathOperator{\diag}{diag}

\DeclareMathOperator\arccosh{arccosh}

\newcommand{\intdd}[1]{\ensuremath{\mathrm{d} #1 \hspace{0.5ex}}}

\begin{document}


\title{On-demand thermoelectric generation of equal-spin Cooper pairs}

\newcommand{\aalto}{Department of Applied Physics,
	Aalto University, 00076 Aalto, Finland}

\newcommand{\duisburg}{Theoretische Physik, 
	Universit\"at Duisburg-Essen and CENIDE, D-47048 Duisburg, Germany}

\newcommand{\wuerzburg}{Institute for Theoretical Physics and Astrophysics,
	University of W\"{u}rzburg, D-97074 W\"{u}rzburg, Germany}
	
\newcommand{\ctqmat}{W\"{u}rzburg-Dresden Cluster of Excellence ct.qmat, Germany}

\author{Felix Keidel}
\affiliation{\wuerzburg}

\author{Sun-Yong Hwang}
\affiliation{\duisburg}

\author{Bj\"orn Trauzettel}
\affiliation{\wuerzburg}
\affiliation{\ctqmat}

\author{Bj\"orn Sothmann}
\affiliation{\duisburg}

\author{Pablo Burset}
\affiliation{\aalto}

\date{\today}


\begin{abstract}
Superconducting spintronics is based on the creation of spin-triplet Cooper pairs in ferromagnet-superconductor (F-S) hybrid junctions. 
Previous proposals to manipulate spin-polarized supercurrents on-demand typically require the ability to carefully control magnetic materials. 
We, instead, propose a quantum heat engine that generates equal-spin Cooper pairs and drives supercurrents on-demand without manipulating magnetic components. 
We consider a S-F-S junction, connecting two leads at different temperatures, on top of the helical edge of a two-dimensional topological insulator. 
Heat and charge currents generated by the thermal bias are caused by different transport processes, where electron cotunneling is responsible for the heat flow to the cold lead and, strikingly, only crossed Andreev reflections contribute to the charge current. 
Such a purely nonlocal Andreev thermoelectric effect injects spin-polarized Cooper pairs at the superconductors, generating a supercurrent that can be switched on/off by tuning their relative phase. 
We further demonstrate that signatures of spin-triplet pairing are facilitated by rather low fluctuations of the thermoelectric current for temperature gradients smaller than the superconducting gap. 
\end{abstract}

\maketitle

%
%
%

%
%
{\it Introduction. ---} 
The new field of \textit{superconducting spintronics} has emerged since the creation of spin-triplet Cooper pairs in experiments~\cite{Eschrig_2010,Linder2015,Eschrig_RPP}. 
The development of spintronics had already benefited from the use of superconducting materials, resulting in longer spin lifetimes and energy-efficient components~\cite{Yang2010,Hubler2012}. 
Now, triplet supercurrents formed by spin-polarized Cooper pairs add the possibility of transporting a net spin component at zero resistance and thus pave the way for spintronic devices that are less liable to overheat~\cite{Sosnin2006,Keizer2006,Robinson2010,Robinson2010a,Khaire2010,Klose2012,Robinson2012,Robinson2014,Srivastava2017,Diesch2018,Jeon2018}. 
The key challenge in the field is the nonequilibrium and on-demand generation of equal-spin Cooper pairs in a viable fashion~\cite{Banerjee2014,Bathen2017,Breunig_2018,Ouassou2018,He2019}, desirably avoiding the complicated manipulation of magnetic components. 

\begin{figure}[b]
    \includegraphics[scale=0.99]{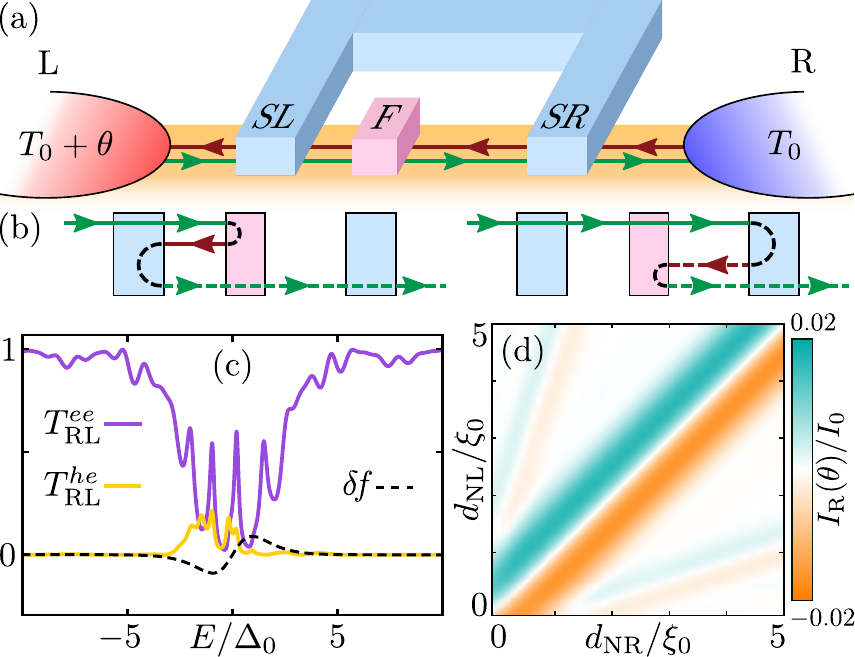}
	\caption{\label{fig:setup} 
		Quantum heat engine generating equal-spin Cooper pairs. 
		(a) SFS junction on the helical edge of a QSHI connecting hot (L) and cold (R) baths. 
		(b) Lowest order contributions to equal-spin CAR. 
		(c) Transmission probabilities for electrons ($T^{ee}_{\mathrm{R} \mathrm{L}}$) and holes ($T^{he}_{\mathrm{R} \mathrm{L}}$), and difference in Fermi distributions $\delta\!f$ with $\theta\!=\!T_c/4$ at $T_0 \! = \! T_c/2$. 
		(d) Unequal distance between F and SL ($d_\text{NL}$) and SR ($d_\text{NR}$) results in an Andreev-dominated thermoelectric current $I_\text{R}(\theta)$. 
	}
\end{figure}

In this Letter, we propose a thermoelectric engine that creates spin-polarized Cooper pairs driving a supercurrent on demand from a temperature gradient. 
We consider a superconductor--ferromagnetic-insulator--superconductor (S-F-S) junction on top of the helical edge state of a \gls{qshi}~\cite{QSHI_2007,Brune_2010,Sullivan_2011,Brune_2012,Reis2017,Kammhuber2017,Wu2018} connecting a hot and a cold bath, \cf \cref{fig:setup}(a). 
Only two microscopic transport processes couple the baths: quantum tunneling of electrons, known as \gls{ec}, and \gls{car}. 
The \gls{qshi} edge states comprise one-dimensional Dirac fermions characterized by spin-momentum locking~\cite{Wu2006,Xu2006}. 
Therefore, while \gls{ec} amounts to a spin-polarized normal current, the peculiar transport properties of the helical edge states guarantee that \gls{car} always converts electrons into holes with the same spin, creating equal-spin Cooper pairs at the superconductors~\cite{Adroguer_2010,Crepin_2015,Keidel_2018}. 
Our key finding is that the nonlocal current becomes dominated by \gls{car} processes, with almost complete suppression of the \gls{ec} contribution. 
This is only possible due to a unique interference effect for \gls{car} processes in our setup. 
As sketched in \cref{fig:setup}(b), \gls{car} requires a spin-flip process at the central ferromagnet and an Andreev reflection at either the left or the right superconductor. 
In an asymmetric junction, the different phases acquired in each path constitute interference, making \gls{car} transmission strongly asymmetric in energy and thus creating an Andreev-dominated thermoelectric current in the right lead, \cf \cref{fig:setup}(c,d). 

Harvesting waste heat by quantum thermoelectric effects has become essential in modern nanoscale devices~\cite{Roche2015}. 
While tackling this problem in S-F hybrid junctions can lead to potentially strong thermoelectric effects~\cite{Machon2013,Kalenkov2014,Kalenkov2015,Bergeret_2014,hwang_large_2016,Kolenda2016,Beiranvand2017,thermoCPS,Hwang2017a,Bergeret_RMP2}, it requires a careful control of magnetic elements and usually features a low heat-to-supercurrent conversion. 
In this proposal, the unique interference of \gls{car} processes, together with the spin-polarization induced by the helical edge state, creates a strong spin-polarized thermoelectric current that can be controlled by tuning the phase difference between the superconducting leads and does not rely on manipulating the ferromagnetic domain. 
We further demonstrate that the thermoelectric current is enhanced over its fluctuations for temperature gradients comparable to the superconducting gap, facilitating the experimental realization of our proposal by thermovoltage~\cite{Kolenda2016,Shelly_2016} or thermophase measurements across the junction~\cite{Bergeret_2015,Giazotto2015}. 

{\it Setup. ---} 
The spin polarization of nonlocal transport and the absence of backscattering at the helical edge of a \gls{qshi} is of great interest for traditional spintronics. 
Moreover, proximity-induced superconductivity and ferromagnetism can confine the helical edge states, opening new scattering channels~\cite{Adroguer_2010,Crepin_2015,Keidel_2018} that can lead to the emergence of Majorana bound states~\cite{FuKane_2009,Crepin_2014,Crepin_2014b,Fleckenstein2018d} or exotic odd-frequency superconducting pairing~\cite{Crepin_2015,Cayao_2017,Keidel_2018,Fleckenstein2018c}. 
Given recent advances in the experimental realization of helical edge states~\cite{Reis2017,Kammhuber2017,Wu2018}, hybrid structures like the one sketched in \cref{fig:setup}(a) are within reach: superconductors~\cite{Sullivan_2012,Hart_2014,Sajadi2018a,Fatemi2018} have been successfully coupled to \glspl{qshi}~\cite{Bocquillon_2016,Bocquillon_2017}, and monolayer \glspl{qshi} provide a new promising platform to induce ferromagnetic order~\cite{Wu2018,Fatemi2018}. 
The observation of Majorana modes in helical hinge states of Bi(111) films under the influence of superconductivity and magnetic iron clusters has recently been reported in Ref.~\cite{Jack2019}. 

We theoretically describe the one-dimensional helical edge states of a \gls{qshi} in proximity to superconducting and ferromagnetic order by a Bogoliubov--de Gennes Hamiltonian in the Nambu basis $\Psi(x) = ( \psi_{\up}, \psi_{\dw}, \psi^{\dagger}_{\dw}, -\psi^{\dagger}_{\up} )$ of the form ($\hbar \! = v_\mathrm{F} \! = \! 1$)
\begin{equation} \label{eq:bdghamiltonian}
 H_{\mathrm{BdG}} = H_{0} + H_{\mathrm{S}} + H_{\mathrm{F}} ,
\end{equation} 
with $H_{0} \! = \! \hat{p}_x \tz \sz - \mu \tz \si$ the Hamiltonian of the free helical edge, $H_{\mathrm{S}} \! = \! [\Delta(x) \cos\phi(x) \tx + \Delta(x) \sin\phi(x) \ty] \si $ the proximity-induced superconductivity, and $H_{\mathrm{F}} \! = \! \ti \mathbf{m}(x) \!\cdot\! \bm{\sigma} \!  \equiv \! \ti  ( m_{\parallel} \cos\lambda \, \sx + m_{\parallel} \sin\lambda \, \sy + m_{z} \sz)$ describing the effect of the ferromagnetic barrier. Here, $\hat{p}_x \!=\! - \ii \partial_x$ and $\hat{\sigma}_i$ ($\hat{\tau}_i$) are Pauli matrices acting in spin (Nambu) space. 
We consider a system with two S regions (named SL and SR) separated by two normal regions (NL and NR) surrounding one ferromagnetic insulator (F); their respective widths are $d_{\mathrm{X}}$ for $\mathrm{X} \!\in\! \{\mathrm{SL,NL,F,NR,SR}\}$. 
The pair potential is assumed equal for both superconductors and constant, $\Delta(x)\!=\!\Delta_0$, a valid approximation as long as the Fermi wavelength in each superconductor is much smaller than the proximity-induced coherence length $\xi_0\!=\!1/\Delta_0$. For simplicity, we take the phase of the pair potential $\phi(x)\!=\!\phi$ in SR and zero otherwise. 
The F region is modeled by constant $m_{\parallel}(x)\!=\!m_0$ within F, and we choose $m_{z}\!=\!0$ since its effect can be absorbed in the phase difference $\phi$ between the superconductors~\cite{Nilsson2008,Crepin_2014,Keidel_2018}. 
Without loss of generality the angle $\lambda$ is set to zero. Finally, we assume that all regions reside at the same chemical potential, \ie $\mu(x)\!=\!0$ everywhere.

In the following, we consider that all leads except $\mathrm{L}$ are at the same temperature~\footnote{To account for finite temperatures in the superconductors, the gap is approximated as $\Delta(T)\!\simeq\!\Delta_0 \tanh(1.74 \sqrt{T_{c}/T-1})$, with $T_{c}$ the critical temperature and $T$ the superconductor's temperature.} ($T_{\mathrm{SL}}\!=\!T_{\mathrm{SR}}\!=\!T_{\mathrm{R}}\!\equiv\!T_0$) and set $T_{\mathrm{L}} \!=\! T_{0} + \theta$, introducing the temperature difference $\theta$. 
The electric current in the right lead after a temperature bias is applied to the left lead is given by $I_{\mathrm{R}} \!=\! I_{\mathrm{R}}^{he} + I_{\mathrm{R}}^{ee}$, where~\footnote{See Supplemental Material, which includes References~\cite{Anantram-Datta,Imry_1986,Bardas_1995,Linder_2016,Butcher_1990,Lumbroso2018,Sigrist_RMP,Legget_RMP}, at [URL will be inserted by publisher] for details of the derivation of the nonlocal current and the supercurrent, an analysis of the \gls{car} interference, the calculation of the anomalous Green function, and more information on the current fluctuations in the right lead.}
\begin{subequations} \label{eq:nonlocalcurrent}
\begin{align} \label{eq:nonlocalcurrentandreev}
I_{\mathrm{R}}^{he} &= I_{0} \int_{-\infty}^{\infty} \frac{\mathrm{d} E}{\Delta_0} \, T^{he}_{\mathrm{R} \mathrm{L}}(E) \, \delta \! f(E) , 
\\
I_{\mathrm{R}}^{ee} &= - I_{0} \int_{-\infty}^{\infty} \frac{\mathrm{d} E}{\Delta_0} \, T^{ee}_{\mathrm{R} \mathrm{L}}(E) \, \delta \! f(E) ,  \label{eq:nonlocalcurrentnormal}
\end{align}
\end{subequations}
with $T^{he}_{\mathrm{R} \mathrm{L}}$ the \gls{car} probability, $T^{ee}_{\mathrm{R} \mathrm{L}}$ the \gls{ec} probability, $I_0 \!=\! e\Delta_0/h$, $\delta \! f(E)\!=\!f[E,0,k_{\mathrm{B}} (T_{0} + \theta)] - f(E,0,k_{\mathrm{B}} T_{0})$, and $f(E,\mu,\tau)\!=\!\{ 1 + \exp[(E - \mu)/\tau]\}^{-1}$ the Fermi distribution function. 
The probabilities are obtained by solving the scattering problem defined by the solutions of \cref{eq:bdghamiltonian} in every region~\cite{BTK,Buttiker_1992,Takane1992,Anantram-Datta,Nazarov2009,Mello2004,Crepin_2014,Crepin_2015,Keidel_2018}. 
Similarly, the current at each superconductor $\text{S}\!=\!\mathrm{SL,SR}$ is given by $I_\text{S}\!=\!\langle J_\text{S}\rangle \!+\! \int_{0}^{d_{\text{S}}} \mathrm{d}x \langle S_\text{S} \rangle$, where $J_\text{S}\!=\!e(\psi^\dagger_{\text{S}\up}\psi_{\text{S}\up}-\psi^\dagger_{\text{S}\dw}\psi_{\text{S}\dw})$ is the quasiparticle current operator and $S_\text{S}\!=\!\ii e\Delta_0(\ee^{-\ii\phi_\text{S}} \psi^\dagger_{\text{S}\up}\psi^\dagger_{\text{S}\dw}+\text{h.c.})$ the source term operator~\cite{Note2}. 

\begin{figure}[t]
	\includegraphics[scale=0.99]{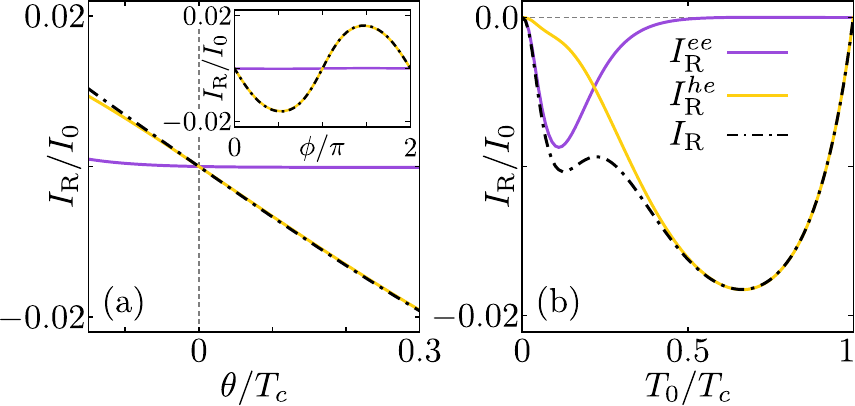}
	\caption{\label{fig:nonlocalcurrent}
		Thermoelectric effect through the S-F-S junction. Total current as well as normal and Andreev contributions as a function of (a) the temperature difference (phase difference $\phi$ in the inset) and (b) the base temperature with $\theta \!=\! T_{0}/2$ fixed. 
		We use the parameters $d_{\mathrm{SL}} \!=\! 
		d_{\mathrm{SR}} \!=\! \xi_0$, $d_{\mathrm{FM}} \!=\! 0.6 \xi_0$, $d_{\mathrm{NL}} \!=\! 0.4 \xi_0$, $d_{\mathrm{NR}} \!=\! 
		0.9 \xi_0$, $m_{0} \!=\! 1.5 \Delta_{0}$, $T_{0} \!=\! 0.5 T_c $, $\phi\!=\!\pi/2$, and $T_c \!=\! \Delta_{0}$ unless specified otherwise. 
	}
\end{figure}

As we describe in detail below, an interference of \gls{car} processes depicted in \cref{fig:setup}(b) leads to a particular thermoelectric effect, where the current can be completely dominated by equal-spin Andreev processes, see \cref{fig:nonlocalcurrent}. 
At the same time, the energy current is only given by the symmetric part of the transmissions; therefore, it can be dominated by \gls{ec} processes. 
Such a decoupling of transport processes for the heat and charge currents is a special feature of this setup. 

%
%
%


%
%
{\it Generation of equal-spin Cooper pairs. ---} 
Helicity determines that particles arriving to the right lead have the same spin polarization as the injected particles on the left lead. 
While this does not restrict the quantum tunneling of electrons through the junction (\gls{ec}), \gls{car} processes are only possible if injected electrons and transmitted holes have the same spin~\cite{Adroguer_2010,Crepin_2015,Keidel_2018}. 
By breaking time-reversal symmetry, the $\mathrm{F}$ region facilitates equal-spin \gls{car} processes. 
As sketched in \cref{fig:setup}(b), incident electrons can be transmitted as holes through the junction if at least one spin-flip process takes place at the $\mathrm{F}$ region and one Andreev reflection occurs at either superconductor. 
Crucially, scattering events involving an Andreev reflection at the right superconductor will acquire an extra phase $\phi$ and a phase shift $d_\mathrm{NR}E$ compared to the ones where the reflection takes place at $\mathrm{SL}$, which are only shifted by $d_\mathrm{NL}E$ (we measure $d_\mathrm{NL,NR}$ in units of $\xi_0$). 
The interference between these two processes is a unique property of \gls{car}, not present in \gls{ec}, resulting in an unusually strong asymmetry of the transmission probability with the energy, \cf \cref{fig:setup}(c). 

\gls{car} processes thus require proximity-induced equal-spin pairing~\cite{Crepin_2015,Keidel_2018}, which we analyze by computing the retarded Green function associated to \cref{eq:bdghamiltonian}~\cite{Crepin_2015,Breunig_2018,Keidel_2018}. 
In the basis defined above, the anomalous part of the retarded Green function is written as $\hat{G}^\mathrm{R}_\mathrm{eh}(x,x',E)\!=\! f_0^\mathrm{R}(x,x',E)\si+\mathbf{f}^\mathrm{R}(x,x',E)\cdot\mathbf{\hat{\sigma}}$, where $f_0^\mathrm{R}$ is the singlet and $\mathbf{f}^\mathrm{R}$ the vector of triplet amplitudes~\cite{Note2}. 
To quantify the net spin carried by a Cooper pair, we define the polarization vector $\mathbf{p}(x,x',E)\!=\!\ii\mathbf{f}^\mathrm{R}\times \mathbf{f}^{\mathrm{R}*}$. 
Note that a finite triplet amplitude is not enough to obtain a polarization. That is the case without magnetic impurity $\mathrm{F}$, where we have $\mathbf{f}^\mathrm{R}\!=\!(0,0,f^\mathrm{R}_3)$, but $\mathbf{p}\!=\!0$. 
By contrast, when time-reversal symmetry is broken by the $\mathrm{F}$ region ($m_0\!\neq\!0$), we find a finite axial polarization $\mathbf{p}(x,x,E)\!=\!(0,p_2,0)$ evidencing that the Cooper pairs develop a net spin. 

\begin{figure*}[ht]
	\includegraphics[scale=0.99]{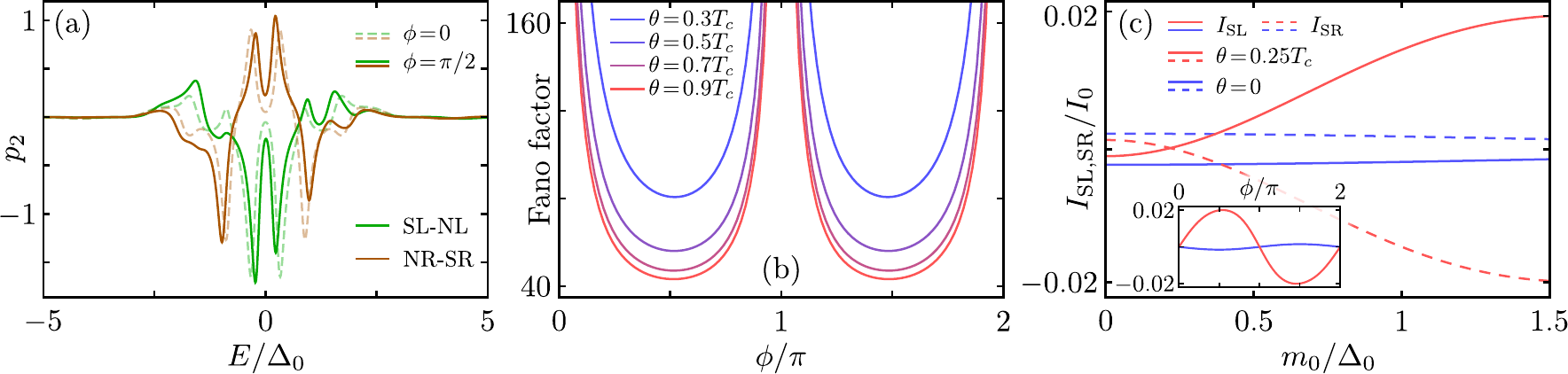}
	\caption{\label{fig:noiseproperties}
		(a) Cooper pair polarization $p_2$, computed at the SL-NL and NR-SR interfaces, as a function of the energy for fixed phase difference $\phi \!=\! \pi/2$ ($\phi \!=\! 0$ for dashed lines). 
		(b) Fano factor for selected temperature differences. 
		(c) Supercurrent in SL (solid lines) and SR (dashed lines) as a function of the magnetic field strength (inset: phase difference), calculated at equilibrium (blue) and for a thermal bias $\theta \!=\! T_c/4$ (red). 
		Rest of parameters are the same as in \cref{fig:nonlocalcurrent}. 
	}
\end{figure*}

We show in \cref{fig:noiseproperties}(a) the polarization of Cooper pairs computed at the interface between each superconductor and the inner normal regions, see
\cref{fig:setup}. 
First, polarizations at each superconductor have different signs, indicating that Cooper pairs with opposite net spin have been transferred to each superconductor. Second, the polarization is maximum for the resonant energies of the S-F-S junction. 
Resonant scattering at each S-F region always gives rise to zero-energy Majorana (quasi-)bound states, with additional finite energy Andreev states depending on the cavity's width~\cite{FuKane_2009,Crepin_2014,Crepin_2015,Keidel_2018,Fleckenstein2018d}. 
The hybridization between the bound states at each S-F cavity is controlled by the phase difference between the superconductors~\cite{Nilsson2008,Keidel_2018}. This, in turn, allows for the control of the electric current through the junction.

%
%
%


%
%
{\it Andreev-dominated thermoelectric effect. ---} 
Given a positive temperature gradient, we find that a finite thermoelectric current, $I_\mathrm{R}$, is completely dominated by Andreev processes when three requirements are fulfilled (see \cref{fig:nonlocalcurrent}): 
\begin{enumerate*}[label=(\roman*)]
 \item the base temperature $T_{0}$ is sufficiently large, \ie $T_{0} \gtrsim T_{c}/2$; 
 \item the junction is asymmetric, which we realize by setting $d_{\mathrm{NL}} \neq d_{\mathrm{NR}}$; and
 \item the phase difference $\phi$ is not an integer multiple of $\pi$.
\end{enumerate*}

Under these conditions, the energy asymmetry of the \gls{car} transmission is comparable to the energy-antisymmetric bias $\delta\!f$ as illustrated in \cref{fig:setup}(c), whereas the asymmetry in the \gls{ec} probability occurs on a much smaller energy scale~\footnote{The physical origin of the asymmetry of \gls{ec} is the spin-splitting of bound states, and it is thus of the order of the hybridization energy. By contrast, the asymmetry in \gls{car} is the result of an interference effect. Note that the location of the extremes of $\delta\!f$, and thus the width of the integration window in \cref{eq:nonlocalcurrent}, are mainly determined by the base temperature $T_0$.}.
As a result, the Andreev current $I_{\mathrm{R}}^{he}$ becomes much larger than the \gls{ec} current $I_{\mathrm{R}}^{ee}$ as the temperature grows. 
The \gls{car} contribution is suppressed as the base temperature approaches $T_{c}$, where the induced gap vanishes. 
It is a good consistency check that simultaneously $I_{\mathrm{R}}^{ee} \!\rightarrow\! 0$, since without superconductivity the resonant tunneling at the S-F regions disappears and so does the thermoelectric effect~\cite{[{Here, we discuss the case of a ferromagnetic insulator, as opposed to another recent proposal to use the edge states of a quantum spin Hall insulator for efficient thermoelectricity, }] Gresta2019}.

The interference effect of \gls{car} processes is caused by an asymmetric S-F-S junction. When $d_{\mathrm{NL}} \!=\! d_{\mathrm{NR}}$, $I_{\mathrm{R}}^{he}$ vanishes since the two paths in \cref{fig:setup}(b) destructively interfere. 
In general, these two contributions acquire a different energy-dependent phase coming from the fact that the Andreev reflection for each path takes place at different superconductors. Consequently, a \gls{car}-dominated thermoelectric current requires that the S-F-S junction is asymmetric and there is a finite phase difference between superconductors, as shown in \cref{fig:setup}(d). 
The interference effect on the \gls{car} probability is written as~\cite{Note2} 
\begin{equation}\label{eq:CARproperties}
 T^{he}_{\mathrm{RL}}(E,\phi) = \gamma(E) \, \cos^{2} \left[ \phi/2 + \left( d_{\mathrm{NR}} - d_{\mathrm{NL}} \right) E \right] ,
\end{equation}
where $\gamma(E)$ is an even function of the energy and $\phi$ is the phase acquired by Andreev reflections at $\mathrm{SR}$. 
Importantly, all higher order contributions are equal for both paths and even in energy~\cite{Note2}, so they are included into the parameter $\gamma(E)$. 
Since only the odd part of $T_{\mathrm{RL}}^{he}$ contributes to the integration, it can be more conveniently expressed as 
\begin{equation} \label{eq:oddCAR}
 T_{\mathrm{RL}}^{he}(-E) - T_{\mathrm{RL}}^{he}(E)  
  = \gamma(E) \sin[2 E (d_{\mathrm{NR}} - d_{\mathrm{NL}})] \sin\phi .
\end{equation}
The sinusoidal behaviour of the current with $\phi$ is shown in the inset of \cref{fig:nonlocalcurrent}(a), revealing the phase difference as an ideal knob to tune the thermoelectric effect. 
\cref{eq:oddCAR} clearly displays two of the three conditions for the Andreev-dominated thermoelectric effect. 
A finite electric current, $I_\mathrm{R}$, is obtained when the phase difference and the asymmetry result in a finite contribution to \cref{eq:oddCAR} that is comparable to the integration window determined by the temperature bias $\delta \! f(T_0,\theta)$. 
We also note that the finite thermoelectric effect indicates the simultaneous presence of both even- and odd-frequency pairing amplitudes in our setup~\cite{Hwang2017a}. 

{\it Signatures of equal-spin Cooper pairs. ---} 
An Andreev-dominated thermoelectric effect stems from the injection of Cooper pairs with opposite net spin into each superconductor. The resulting current, $I_{\mathrm{R}}$, could be measured as a voltage drop on the right lead after heating up the left one~\cite{Kolenda2016,Karimi_2018}. Our proposal is also phase-tuneable so the thermoelectric power generation can be controlled by a small magnetic field~\cite{Shelly_2016}. 
Increasing the temperature gradient drives larger thermoelectric currents [see \cref{fig:nonlocalcurrent} (a)], but also potentially larger fluctuations~\cite{Kheradsoud2019}. It is thus essential for the characterization of the proposed heat engine to identify a parameter regime where the fluctuations are the smallest with respect to the average current. That is, where the Fano factor $F = S_{\mathrm{RR}}/\abs{2e I_{\mathrm{R}}}$, with $S_{\mathrm{RR}}$ the current fluctuations in the right lead, is minimal. 
The zero-frequency fluctuation of $I_{\mathrm{R}}$ is given by~\cite{Anantram-Datta}
\begin{multline} \label{eq:noiseanantramdatta}
 S_{\mathrm{RR}} = e I_{0} \int_{-\infty}^{\infty} \frac{\mathrm{d} E}{\Delta_0} \, \sum_{\alpha,\beta,\gamma,\delta,k,l}  \sgn(\alpha) \sgn(\beta)    \\
 \times 
 A_{k \gamma, l \delta}(\mathrm{R} \alpha, E) A_{l \delta, k \gamma}(\mathrm{R} \beta, E) \, 
 f_{k \gamma}(E)  \, \left[ 1 - f_{l \delta}(E)\right] ,
\end{multline}
with 
 \begin{equation}
  A_{k \gamma, l \delta}(i \alpha, E) = \delta_{i, k} \delta_{i, l} \delta_{\alpha, \gamma} \delta_{\alpha, \delta} - \left[ s_{i k}^{\alpha \gamma}(E) \right]^{*} s_{i l}^{\alpha \delta}(E) ,
 \end{equation} 
where Greek letters label Nambu indices, with $\sgn(\alpha)\!=\!\pm1$ for $\alpha\!=\!e,h$, Latin symbols represent reservoirs $\{ \mathrm{L}, \mathrm{R}, \mathrm{SL}, \mathrm{SR}\}$, $s_{ik}^{\alpha\gamma}$ denotes the amplitude for a particle of type $\gamma$ in reservoir $k$ to be scattered into reservoir $i$ as a particle of type $\alpha$, and $f_{j\beta}(E) \!=\! f(E,\sgn(\beta)\mu_{j},k_{\mathrm{B}}T_{j})$ is the Fermi distribution for particles $\beta$ in reservoir $j$~\footnote{Since we neglect quasiparticle injection in the superconductors, the sum over the reservoirs in \cref{eq:noiseanantramdatta} is effectively restricted to L and R.}. 

For an asymmetric junction, the phase difference $\phi$ controls the thermoelectric current, see \cref{eq:oddCAR}. The fluctuations are, however, almost independent of $\phi$, indicating that they are mostly caused by thermal noise~\cite{Note2}. 
Due to the carrier-selective heat and charge transfer in this setup, thermal noise is caused by normal scattering processes that do not experience interference and $S_{\mathrm{RR}}$ increases steadily with the temperature bias $\theta$. 
By contrast, the Andreev-dominated current increases rapidly for $\theta\!<\!T_c$ and appears to saturate for higher bias. 
Importantly, for experimentally relevant values~\cite{Kolenda2016,Karimi_2018}, when the current is maximum, the Fano factor becomes minimum, see \cref{fig:noiseproperties}(b), thus demonstrating that the current is enhanced over its fluctuations. 
Note that an Andreev-dominated current requires $T_0 \!\gtrsim\!T_c/2$, resulting in rather large Fano factors~\cite{Note2}. 
Recently, the electronic noise due to temperature differences in mesoscopic conductors, different than thermal or shot noise, was measured and proposed as an accurate temperature probe~\cite{Lumbroso2018}. 

Finally, the injection of Cooper pairs with opposite net spin into each superconductor produces a non-equilibrium supercurrent $I_{\mathrm{SL}}(\phi)\!=\!I_{\mathrm{SR}}(-\phi)$, see \cref{fig:noiseproperties}(c). At equilibrium ($\theta\!=\!0$), even though the current at the normal leads vanishes, $I_\mathrm{L,R}\!=\!0$, there is a finite, Josephson-like supercurrent at each superconductor. 
Interestingly, without the F region (\ie with $m_0\!=\!0$ and no \gls{car}), the equilibrium current is barely affected by the temperature bias. By contrast, in the presence of F, \gls{car} processes pump equal-spin Cooper pairs into the superconductors resulting in large non-equilibrium supercurrents. 
The temperature bias thus creates supercurrents with opposite sign at each superconductor, that could be measured separately or after connecting them through a loop, as depicted in \cref{fig:setup}. Measuring the flux inside the loop with and without temperature gradient, one could determine a thermophase~\cite{Bergeret_2015}. 
Within our estimations, for biases $\theta\!<\!T_c$, close to the minimum of fluctuations, the magnitude of the temperature-induced supercurrent is comparable to $I_0$, the zero-temperature maximum Josephson current, with a typical value of $\sim\!1\mu$A. 

{\it Summary. ---} 
We propose a quantum heat engine that can be electrically controlled to generate spin-polarized Cooper pairs and drive supercurrents from a temperature bias on demand. 
Our proposal is based on a unique transport mechanism taking place at a S-F-S junction on the helical edge of a \gls{qshi}. Nonlocal Andreev processes through the junction experience an interference effect between the contributions from each superconductor. 
This interference is not present for normal processes, resulting in carrier-selective heat and charge currents where normal processes transfer heat and Andreev processes transfer charge. 
Due to the strong spin-orbit coupling at the helical edge state, the thermoelectric current is completely dominated by equal-spin Andreev processes. 
We discussed how the proposed spin-triplet thermoelectric effect could be measured as a thermophase appearing between the superconductors. The measurement is further facilitated by the low fluctuations of the spin-polarized nonlocal current. 

The authors are grateful to M. Moskalets for valuable discussions. 
We acknowledge support from the DFG (SPP 1666 and SFB 1170, Project-ID 258499086), the Cluster of Excellence EXC 2147 (Project-ID 39085490), the Ministry of Innovation NRW via the ``Programm zur F\"orderung der R\"uckkehr des hochqualifizierten Forschungsnachwuchses aus dem Ausland'', the Horizon 2020 research and innovation programme under the Marie Sk\l odowska-Curie Grant No. 743884 and the Academy of Finland (project 312299).

%
%


\bibliography{TIS_TH}


\clearpage

\onecolumngrid

\setcounter{equation}{0}
\renewcommand{\theequation}{S\,\arabic{equation}}

\setcounter{figure}{0}
\renewcommand{\thefigure}{S\,\arabic{figure}}

\section{Supplemental material to ``On-demand thermoelectric generation of equal-spin Cooper pairs"}

In this supplemental material, we provide more details regarding the derivation of the expression for the current in the right lead and the crossed Andreev reflection amplitude.
We also describe the Green function techniques used to compute the Cooper pair polarization. Finally, we show the behavior of the current fluctuations in the right lead as a function of the base temperature.

\section{Current in the right lead}

In order to derive Eq.~(2) of the main text, we follow the formalism developed in Refs.~\cite{Imry_1986,Bardas_1995,Linder_2016,Butcher_1990}. 
As a starting point, by virtue of the standard approach in mesoscopic physics~\cite{Anantram-Datta} and to recapitulate the main text, the charge current on the right side of the setup $I_{\mathrm{R}}$ is given by
\begin{equation} \label{supeq:currentdefinition}
 I_{\mathrm{R}} = 
 \frac{e}{2h} \int_{-\infty}^{\infty} \mathrm{d} E \, \sum_{\alpha,\beta,j}  \sgn(\alpha)   \left[ \delta_{ \mathrm{R},j} \delta_{\alpha,\beta} - T_{\mathrm{R}j}^{\alpha \beta}(E) \right] f_{j \beta}(E),
\end{equation}
where Greek summation indices $\alpha,\beta \in \{ e,h \}$ run over the electron/hole degree of freedom with $\sgn(\alpha)=\pm1$ for $\alpha=e/h$, the Latin index $j \in \{ \mathrm{L}, \mathrm{R}, \mathrm{SL}, \mathrm{SR}\}$ runs over all reservoirs, $T_{Rj}^{\alpha\beta}\!=\!|s_{Rj}^{\alpha\beta}|^2$, with $s_{Rj}^{\alpha\beta}$ the scattering amplitude for a particle of type $\beta$ in reservoir $j$ to be scattered into reservoir $\mathrm{R}$ as a particle of type $\alpha$, and $f_{j\beta}(E) = f(E,\sgn(\beta)\mu_{j},k_{\mathrm{B}}T_{j})$, with $f(E,\mu,\tau)=\{ 1 + \exp[(E - \mu)/\tau]\}^{-1}$, is the Fermi distribution for particles $\beta$ in reservoir $j$.

The lengthy expression arising from performing the sum in \cref{supeq:currentdefinition} can be simplified substantially. Using our assumption of grounded superconductors, \ie $\mu_{\mathrm{SL}} = \mu_{\mathrm{SR}} = 0$, and equal superconductor temperatures, \ie $T_{\mathrm{SL}} = T_{\mathrm{SR}} \equiv T_{0}$, the Fermi functions in the superconductors coincide and $f_{\mathrm{SL} e}(E) = f_{\mathrm{SL} h}(E) = f_{\mathrm{SR} e}(E) = f_{\mathrm{SR} h}(E) \equiv f_{0}(E) = f(E,0,k_{\mathrm{B}} T_{0})$ follows. 
Furthermore, by employing unitarity of the scattering matrix and conservation of quasiparticle current, which implies
\begin{equation}
 \sum_{j,\beta} T_{ij}^{\alpha\beta} = \sum_{i, \alpha} T_{ij}^{\alpha\beta} = 1 ,
\end{equation} 
we can eliminate the coefficients involving $\mathrm{SL}, \mathrm{SR}$ and arrive at
\begin{equation} \label{supeq:fullcurrentbeforephs}
\begin{gathered} 
 I_{\mathrm{R}} = \frac{e}{2h} \int_{-\infty}^{\infty} \mathrm{d} E \, 
 \left\lbrace 
 \left[ 1 - T^{ee}_{\mathrm{R} \mathrm{R}}(E) + T^{he}_{\mathrm{R} \mathrm{R}}(E)  \right] \left[f_{\mathrm{R} e}(E) - f_{0}(E) \right] 
 - \left[ 1 - T^{hh}_{\mathrm{R} \mathrm{R}}(E) + T^{eh}_{\mathrm{R} \mathrm{R}}(E)  \right] \left[f_{\mathrm{R} h}(E) - f_{0}(E) \right] 
 \right.  \\*  \left.
 + \left[ T^{he}_{\mathrm{R} \mathrm{L}}(E) - T^{ee}_{\mathrm{R} \mathrm{L}}(E)  \right] \left[f_{\mathrm{L} e}(E) - f_{0}(E) \right] 
 - \left[ T^{eh}_{\mathrm{R} \mathrm{L}}(E) - T^{hh}_{\mathrm{R} \mathrm{L}}(E)  \right] \left[f_{\mathrm{L} h}(E) - f_{0}(E) \right] 
 \right\rbrace .
\end{gathered} 
\end{equation}

Next, by recognizing that $f_{ih}(E) - f_{0}(E) = f_{0}(-E) - f_{ie}(-E)$ and that particle-hole symmetry enforces $T_{ij}^{\alpha\beta}(E) = T_{ij}^{\bar{\alpha}\bar{\beta}}(-E)$ where $\bar{\alpha} = h,e \text{ if } \alpha = e,h$, the terms in \cref{supeq:fullcurrentbeforephs} corresponding to the injection of holes can be folded back onto their charge conjugated counterparts, yielding
\begin{equation}  \label{supeq:fullcurrent}
 \begin{gathered}
 I_{\mathrm{R}} = \frac{e}{h} \int_{-\infty}^{\infty} \mathrm{d} E \, 
 \left\lbrace 
 \left[ 1 - T^{ee}_{\mathrm{R} \mathrm{R}}(E) + T^{he}_{\mathrm{R} \mathrm{R}}(E)  \right]
 \left[f_{\mathrm{R} e}(E) - f_{0}(E) \right] 
 + \left[ T^{he}_{\mathrm{R} \mathrm{L}}(E) - T^{ee}_{\mathrm{R} \mathrm{L}}(E)  \right] \left[f_{\mathrm{L} e}(E) - f_{0}(E) \right]  \right\rbrace .
\end{gathered} 
\end{equation}
The first term in \cref{supeq:fullcurrent} corresponds to local current contributions in the right lead resulting from injection from the right reservoir, while the second term in \cref{supeq:fullcurrent} describes the charge current in the right lead rooting in \gls{car} and \gls{ec} processes of particles injected from the left reservoir. 

We stress that for our choice of chemical potentials and temperatures \cref{supeq:fullcurrent} is general for a four terminal setup with two superconducting leads, irrespective of the specific scattering problem at hand. 
\cref{supeq:fullcurrent} holds even when including quasiparticle injection from and into the superconductors.

Importantly, since we assume equilibrium between the right reservoir and the superconductors, \ie $T_{\mathrm{R}} = T_{0}$, the local current contribution vanishes identically.
Therefore, the current in the right lead is a purely nonlocal effect and solely given by 
$I_{\mathrm{R}} = I^{he}_{\mathrm{R}} + I^{ee}_{\mathrm{R}}$, with
\begin{subequations} \label{supeq:nonlocalcurrents}
\begin{gather} \label{supeq:nonlocalcurrentandreev}
I^{he}_{\mathrm{R}} = I_{0} \int_{-\infty}^{\infty} \frac{\mathrm{d E}}{\Delta_0} \, T^{he}_{\mathrm{R} \mathrm{L}}(E) \left[f_{\mathrm{L}e}(E) - f_{0}(E) \right] ,
\shortintertext{and} 
I^{ee}_{\mathrm{R}} = - I_{0} \int_{-\infty}^{\infty} \frac{\mathrm{d E}}{\Delta_0} \, T^{ee}_{\mathrm{R} \mathrm{L}}(E) \left[ f_{\mathrm{L}e}(E) - f_{0}(E) \right] ,  \label{supeq:nonlocalcurrentnormal}
\end{gather}
\end{subequations}
where $I_{0} = e\Delta_0/h$.
Upon identifying $f_{\mathrm{L}e}(E) - f_0(E) \equiv \delta \! f(E)$, \cref{supeq:nonlocalcurrentandreev,supeq:nonlocalcurrentnormal} correspond to Eqs.~(2a) and (2b) of the main text.

The helicity of the \gls{qshi} edge states profoundly affects the resulting current. By convention, incoming and right-moving particles from the left \textit{and} outgoing right-moving particles and holes on the right of the system must have spin $\up$. Consequently, the edge states act as a spin filter for nonlocally driven current.

\section{Current in the superconductors}

The current operator in Nambu and spin space is defined as
\begin{equation}
\hat{I}(x) = \frac{1}{2} e v_{\mathrm{F}} \Psi^{\dagger} 
\diag \left( 1, -1, 1, -1 \right)
\Psi ,
\end{equation}
with $\Psi = (\psi_\uparrow, \psi_\downarrow, \psi_\downarrow^{\dagger}, -\psi_\uparrow^{\dagger} )^{T}$. 
The field operators can be rewritten using the scattering states, according to 
\begin{equation}
\Psi(x) = \sum_{i \in \{1,2,3,4 \} } \int_{-\infty}^{\infty} \frac{\intdd{E}}{\sqrt{h v_F}} \phi_i(x) \hat{a}_i ,
\end{equation} 
where the $\phi_i$ are the scattering states and thus solutions to the BdG Hamiltonian, and the sum over the index $i$ runs over the four scattering states ($i=1$: e from left, $i=2$: h from left, $i=3$: e from right, $h$ from right). Here, $\hat{a}_i$ is the electron or hole annihilation operator at the corresponding lead, L or R. 
This means that the index $i$ contains both lead and particle type index from the sum in Eq. (23) of Ref.~\cite{Anantram-Datta}.  
The components of the scattering states are given by $\phi_i(x) = (u_{\uparrow, i}, u_{\downarrow, i}, v_{\downarrow, i}, v_{\uparrow, i}  )^{T}$.
With that, the average current can be written as 
\begin{equation} \label{eq:xdependentcurrent}
\mean{\hat{I}(x)} = \frac{e}{2h} \int_{-\infty}^{\infty} \intdd{E} \sum_i \left\lbrace
\abs{u_{\uparrow, i}(x)}^{2} f_i - \abs{u_{\downarrow, i}(x)}^{2} f_i + \abs{v_{\downarrow, i}(x)}^{2} f_i - \abs{v_{\uparrow, i}(x)}^{2} f_i 
\right\rbrace ,
\end{equation} 
with $f_{1,2} = f_{\mathrm{L}e,\mathrm{L}h}$ and $f_{3,4} = f_{\mathrm{R}e,\mathrm{R}h}$.  
One can check that this gives back the same result as Eq. (29) of Ref.~\cite{Anantram-Datta}, and therefore the expression we used for the current in the normal leads, for $I_\mathrm{L,R}$ when applied to a two-lead setup (for $I_\mathrm{R}$ one needs to insert a minus sign in order to follow the convention of positive currents always flowing out of the reservoirs). 
\cref{eq:xdependentcurrent} now allows us to calculate the (quasiparticle) current at any position using the known scattering states. 

We now follow \cref{eq:xdependentcurrent} and apply it to the definition of the source term in the main text, resulting in
\begin{equation}
\mean{S} = - 2 e \Delta_0 \sum_{i\in\{ 1,2,3,4 \}} \int \intdd{E} 
\Big[ 
\imag( \ee^{-\ii \phi} u_{\up,i}^* v_{\dw,i} ) + \imag( \ee^{-\ii \phi} u_{\dw,i}^* v_{\up,i} )
\Big] f_i .
\end{equation}
After integration, the supercurrent is given by
\begin{equation}
\int_{x_0}^{x} \intdd{x'} \mean{S} = - 2 e \Delta_0  \int_{x_0}^{x} \intdd{x'} \sum_{i\in\{ 1,2,3,4 \}} \int \intdd{E} 
\Big[ 
\imag( \ee^{-\ii \phi} u_{\up,i}^*(x') v_{\dw,i}(x') ) + \imag( \ee^{-\ii \phi} u_{\dw,i}^*(x') v_{\up,i}(x') )
\Big] f_i .
\end{equation}
Note that, to take into account the temperature of the superconductor, $\Delta_0$ should be replaced by $\Delta(T_\mathrm{S})$. 
Furthermore, note that within the superconducting regions, there is also a quasiparticle contribution of the same form as the current in the normal regions in \cref{eq:xdependentcurrent}.

\section{Analysis of the crossed Andreev reflection amplitude}


\begin{figure}[t]
\begin{center}
\includegraphics[scale=0.99]{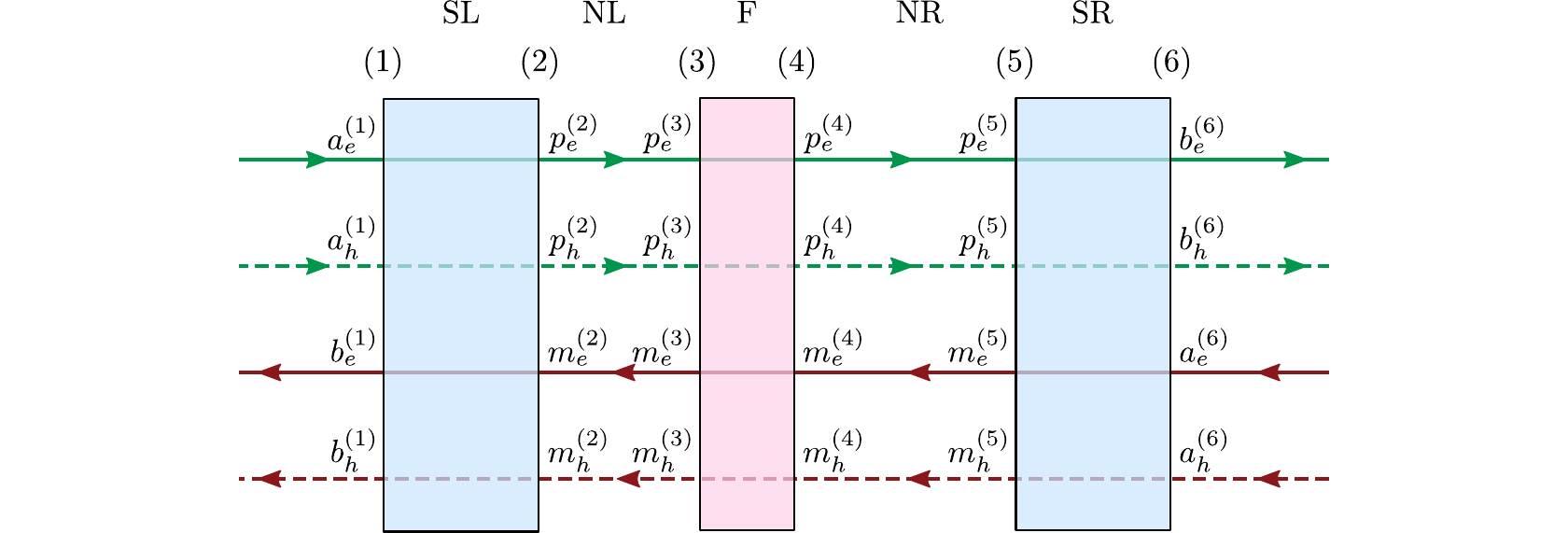}
\end{center}
\caption{\label{supfig:CARdecompsystem} 
Sketch of the scattering region. Full (dashed) lines correspond to electrons (holes), whereas the color distinguishes spin up (green) from spin down (brown) modes. Additionally, the arrows indicate the direction of propagation. The shaded rectangles indicate the barriers, and the interfaces are numbered for clarity.
}
\end{figure}


In this section of the supplementary material, we discuss the \gls{car} amplitude in more detail and provide a derivation of Eqs.~(3) and (4) of the main text.
To that end, we decompose the full scattering problem into simpler pieces, namely the superconducting barriers SL and SR, the ferromagnetic region F, and the intermediate normal domains NL and NR.
We follow the same approximations as stated in the main text. 
For the sake of readability we slightly change notation compared to the main text and denote transmissions and reflections as $t_{ij}^{\alpha\beta}$ with $i \neq j$ and $r_{ii}^{\alpha\beta}$, respectively.

In order to write the S-matrix elements of the full system in terms of the scattering coefficients of the single constituents, we proceed as follows. Globally, the amplitudes of incoming and outgoing modes at the outmost interfaces $(1)$ and $(6)$ (see \cref{supfig:CARdecompsystem}) are related by the full scattering matrix $S$ according to 
\begin{equation} \label{supeq:Sglobal}
\begin{pmatrix}
b_{e}^{(1)} \\[2pt]
b_{h}^{(1)} \\[2pt]
b_{e}^{(6)} \\[2pt]
b_{h}^{(6)}
\end{pmatrix}
=
S \, 
\begin{pmatrix}
a_{e}^{(1)} \\[2pt]
a_{h}^{(1)} \\[2pt]
a_{e}^{(6)} \\[2pt]
a_{h}^{(6)}
\end{pmatrix} ,
\end{equation}
with
\begin{equation}
S= 
\begin{pmatrix}
R && T' \\
T && R'
\end{pmatrix}
\quad \text{and} \quad
R= 
\begin{pmatrix}
r^{ee}_{\mathrm{LL}} && r^{eh}_{\mathrm{LL}} \\[2pt]
r^{he}_{\mathrm{LL}} && r^{hh}_{\mathrm{LL}}
\end{pmatrix} \, , \,
R'= 
\begin{pmatrix}
r^{ee}_{\mathrm{RR}} && r^{eh}_{\mathrm{RR}} \\[2pt]
r^{he}_{\mathrm{RR}} && r^{hh}_{\mathrm{RR}}
\end{pmatrix} \, , \,
T= 
\begin{pmatrix}
t^{ee}_{\mathrm{RL}} && t^{eh}_{\mathrm{RL}} \\[2pt]
t^{he}_{\mathrm{RL}} && t^{hh}_{\mathrm{RL}}
\end{pmatrix}  \, , \,
T'= 
\begin{pmatrix}
t^{ee}_{\mathrm{LR}} && t^{eh}_{\mathrm{LR}} \\[2pt]
t^{he}_{\mathrm{LR}} && t^{hh}_{\mathrm{LR}}
\end{pmatrix}.
\end{equation}
Here, $a^{(i)}_{\alpha}$ ($b^{(i)}_{\alpha}$) with $i=1,6$ denotes the amplitude of an incoming (outgoing) mode at interface $(i)$ of particle/hole type $\alpha$. For the scattering coefficients of the total system we choose the convention that the right (left) sub- and superscript refers to the incoming (outgoing) particle, e.g. $t_{\mathrm{RL}}^{he}$ is the amplitude for an electron to be crossed Andreev reflected from left to right.

Within the system the amplitudes between positions $(i)$ and $(i+1)$ are related by scattering matrices associated with single S and F barriers or the intermediate NL and NR regions. We denote the amplitude of rightmovers (leftmovers) of type $\alpha$ at interface $(i)$ with $p_{\alpha}^{(i)}$ ($m_{\alpha}^{(i)}$). Specifically, they are related by
\begin{subequations} \label{supeq:Smatrices}
\begin{alignat}{4} \label{supeq:SSL}
\begin{pmatrix}
b_{e}^{(1)} \\[2pt]
b_{h}^{(1)} \\[2pt]
p_{e}^{(2)} \\[2pt]
p_{h}^{(2)}
\end{pmatrix}
&=
S_{\mathrm{SL}} \, 
&&
\begin{pmatrix}
a_{e}^{(1)} \\[2pt]
a_{h}^{(1)} \\[2pt]
m_{e}^{(2)} \\[2pt]
m_{h}^{(2)}
\end{pmatrix} 
&& \quad\text{with} \quad &
S_{\mathrm{SL}}
&=
\begin{pmatrix}
0	&&	r^{eh}_{\mathrm{SL}(11)}	&&	t^{ee}_{\mathrm{SL}(12)}	&&	0	\\[2pt]
r^{he}_{\mathrm{SL}(11)}	&&	0	&&	0	&&	t^{hh}_{\mathrm{SL}(12)}	\\[2pt]
t^{ee}_{\mathrm{SL}(21)}	&&	0	&&	0	&&	r^{eh}_{\mathrm{SL}(22)}	\\[2pt]
0	&&	t^{hh}_{\mathrm{SL}(21)}	&&	r^{he}_{\mathrm{SL}(22)}	&&	0
\end{pmatrix} , \\[2pt]
\label{supeq:SNL}
\begin{pmatrix}
m_{e}^{(2)} \\[2pt]
m_{h}^{(2)} \\[2pt]
p_{e}^{(3)} \\[2pt]
p_{h}^{(3)}
\end{pmatrix}
&=
S_{\mathrm{NL}} \, 
&&
\begin{pmatrix}
p_{e}^{(2)} \\[2pt]
p_{h}^{(2)} \\[2pt]
m_{e}^{(3)} \\[2pt]
m_{h}^{(3)}
\end{pmatrix} 
&& \quad\text{with} \quad &
S_{\mathrm{NL}}
&=
\begin{pmatrix}
0&&	0 &&	t^{ee}_{\mathrm{NL}(23)}	&&	0\\[2pt]
0&&	0 &&	0	&&	t^{hh}_{\mathrm{NL}(23)}\\[2pt]
t^{ee}_{\mathrm{NL}(32)}	&&	0 &&	0 &&	0\\[2pt]
0 &&	t^{hh}_{\mathrm{NL}(32)}	&&	0 && 0
\end{pmatrix} , \\[2pt]
\label{supeq:SF}
\begin{pmatrix}
m_{e}^{(3)} \\[2pt]
m_{h}^{(3)} \\[2pt]
p_{e}^{(4)} \\[2pt]
p_{h}^{(4)}
\end{pmatrix}
&=
S_{\mathrm{F}} \, 
&&
\begin{pmatrix}
p_{e}^{(3)} \\[2pt]
p_{h}^{(3)} \\[2pt]
m_{e}^{(4)} \\[2pt]
m_{h}^{(4)}
\end{pmatrix} 
&& \quad\text{with} \quad &
S_{\mathrm{F}}
&=
\begin{pmatrix}
r^{ee}_{\mathrm{F}(33)}		&&	0	&&	t^{ee}_{\mathrm{F}(34)}	&&	0		\\[2pt]
0	&&	r^{hh}_{\mathrm{F}(33)}		&&	0	&&	t^{hh}_{\mathrm{F}(34)}		\\[2pt]
t^{ee}_{\mathrm{F}(43)}		&&	0	&&	r^{ee}_{\mathrm{F}(44)}	&&	0		\\[2pt]
0	&&	t^{hh}_{\mathrm{F}(43)}		&&	0	&&	r^{hh}_{\mathrm{F}(44)}
\end{pmatrix} , \\[2pt]
\label{supeq:SNR}
\begin{pmatrix}
m_{e}^{(4)} \\[2pt]
m_{h}^{(4)} \\[2pt]
p_{e}^{(5)} \\[2pt]
p_{h}^{(5)}
\end{pmatrix}
&=
S_{\mathrm{NR}} \, 
&&
\begin{pmatrix}
p_{e}^{(4)} \\[2pt]
p_{h}^{(4)} \\[2pt]
m_{e}^{(5)} \\[2pt]
m_{h}^{(5)}
\end{pmatrix} 
&& \quad\text{with} \quad &
S_{\mathrm{NR}}
&=
\begin{pmatrix}
0&&	0&&	t^{ee}_{\mathrm{NR}(45)}	&&0	\\[2pt]
0&&	0&&	0	&&	t^{hh}_{\mathrm{NR}(45)}\\[2pt]
t^{ee}_{\mathrm{NR}(54)}	&&0	&&0	&&0	\\[2pt]
0&&	t^{hh}_{\mathrm{NR}(54)}	&&0	&&0
\end{pmatrix} , \\[2pt]
\label{supeq:SSR}
\begin{pmatrix}
m_{e}^{(5)} \\[2pt]
m_{h}^{(5)} \\[2pt]
b_{e}^{(6)} \\[2pt]
b_{h}^{(6)}
\end{pmatrix}
&=
S_{\mathrm{SR}} \, 
&&
\begin{pmatrix}
p_{e}^{(5)} \\[2pt]
p_{h}^{(5)} \\[2pt]
a_{e}^{(6)} \\[2pt]
a_{h}^{(6)}
\end{pmatrix} 
&& \quad\text{with} \quad &
S_{\mathrm{SR}}
&=
\begin{pmatrix}
0&&	r^{eh}_{\mathrm{SR}(55)}	&&	t^{ee}_{\mathrm{SR}(56)}	&&	0	\\[2pt]
r^{he}_{\mathrm{SR}(55)}	&&	0	&&	0	&&	t^{hh}_{\mathrm{SR}(56)}	\\[2pt]
t^{ee}_{\mathrm{SR}(65)}	&&	0	&&	0	&&	r^{eh}_{\mathrm{SR}(66)}	\\[2pt]
0&&	t^{hh}_{\mathrm{SR}(65)}	&&	r^{he}_{\mathrm{SR}(66)}	&&0
\end{pmatrix} .
\end{alignat}
\end{subequations}
In \cref{supeq:SSL,supeq:SNL,supeq:SF,supeq:SNR,supeq:SSR}, $r^{\alpha\beta}_{X(ii)}$ corresponds to a reflection process of a particle of type $\beta$ into type $\alpha$ at interface $(ii)$ of region X, whereas $t^{\alpha\alpha}_{X(ij)}$ represents a transmission of particle $\alpha$ from interface $j$ to $i$ through region X with $X \in \{ \mathrm{SL},\mathrm{NL},\mathrm{F},\mathrm{NR},\mathrm{SR} \}$.

The scattering problems need to be set up such that the scattering coefficients alone capture the phase shifts picked up due to propagation. 
Specifically, there are four solutions of the Bogoliubov--de Gennes Hamiltonian at each interface $(i)$ given by (note that we set $\hbar = v_\mathrm{F} = 1$)
\begin{equation}
    \phi_{+}^{e}(x) = \ee^{\ii E x} 
    \begin{pmatrix}
    1 \\ 0 \\ 0 \\ 0
    \end{pmatrix} , \quad 
    \phi_{-}^{e}(x) = \ee^{- \ii E x} 
    \begin{pmatrix}
    0 \\ 1 \\ 0 \\ 0
    \end{pmatrix} , \quad 
    \phi_{-}^{h}(x) = \ee^{- \ii E x} 
    \begin{pmatrix}
     0 \\ 0 \\ 1 \\ 0
    \end{pmatrix} , \quad 
    \phi_{+}^{h}(x) = \ee^{\ii E x} 
    \begin{pmatrix}
    0 \\ 0 \\ 0 \\ 1
    \end{pmatrix} ,
\end{equation}
corresponding to right- ($+$) and leftmoving ($-$) electrons and holes ($e$ and $h$, respectively). 
The scattering matrix relating interfaces $(i)$ and $(i+1)$ is then obtained by constructing scattering states out of the solutions $\phi^{\alpha}_{\pm}(x-x_{(i)})$ and $\phi^{\alpha}_{\pm}(x-x_{(i+1)})$, respectively, where $x_(j)$ denotes the location of interface $(j)$.

One can now use all 16 subequations of \cref{supeq:SSL,supeq:SNL,supeq:SF,supeq:SNR,supeq:SSR} not involving the outgoing amplitudes $b^{(i)}_{\alpha}$ in order to write the 16 coefficients $p^{(i)}_{\alpha}$, $m^{(i)}_{\alpha}$ in terms of the ingoing amplitudes $a^{(i)}_{\alpha}$. 
Subsequently, the remaining 4 linear equations can be used to relate the outgoing and ingoing amplitudes $b^{(i)}_{\alpha}$ and $a^{(i)}_{\alpha}$ in the form of \cref{supeq:Sglobal} and we can read off $t^{he}_{\mathrm{RL}}$.


\begin{figure}
 \begin{center}
  \includegraphics[scale=0.99]{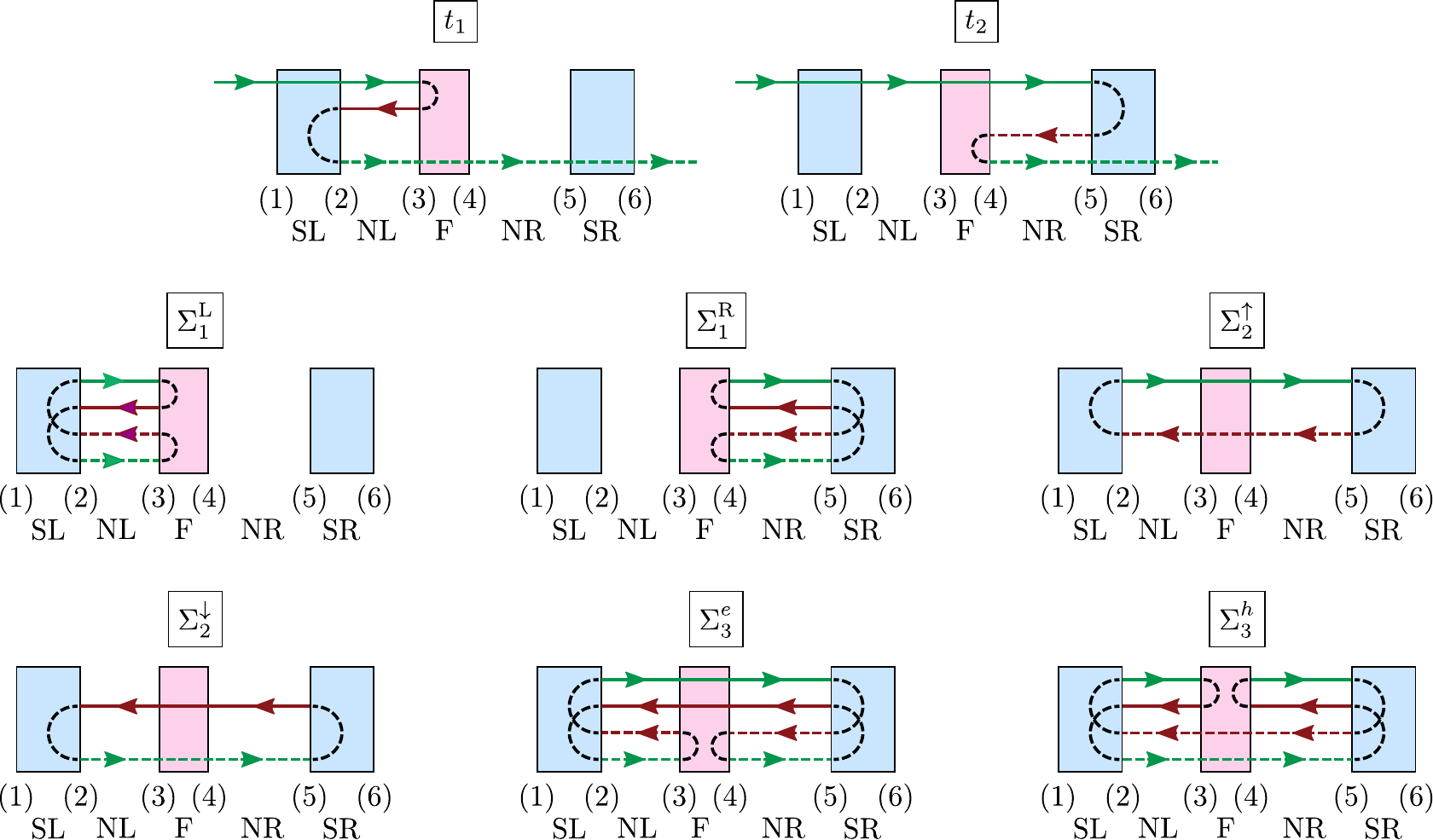}
  \caption{\label{supfig:paths}
  Schematic representation of the contributions to the crossed Andreev reflection amplitude $t_{\mathrm{RL}}^{he}$. Colors and dashing distinguish spin and particle type, as defined in \cref{supfig:CARdecompsystem}. The black dashed lines are added to indicate reflection processes, whereas transmissions are simply represented by lines crossing a scattering region.
  The two lowest order paths $t_{1,2}$ are augmented by the closed loops shown in the lower two rows according to \cref{supeq:caramplitudedecomposed}.
  }
 \end{center}
\end{figure}


We obtain
\begin{equation} \label{supeq:caramplitudedecomposed}
  t^{he}_{\mathrm{RL}} = \frac{t_{1} + t_{2}}{1 - \bm{\varSigma}} = \left( t_{1} + t_{2} \right) \, \sum_{n=0}^{\infty}  \bm{\varSigma}^{n} ,
\end{equation}
where we define self-energies as
\begin{equation}
 \bm{\varSigma} = 
  \Sigma_{1}^{\mathrm{L}} + \Sigma_{1}^{\mathrm{R}} - \Sigma_{1}^{\mathrm{L}} \Sigma_{1}^{\mathrm{R}} 
  + \Sigma_{2}^{\up} + \Sigma_{2}^{\dw} - \Sigma_{2}^{\up} \Sigma_{2}^{\dw} 
  +\Sigma_{3}^{e} + \Sigma_{3}^{h} .
\end{equation} 

Before we provide the lengthy expressions for $t_{1}$, $t_{2}$ and $\bm{\varSigma}$ in terms of the elements of the scattering matrices defined in \cref{supeq:Smatrices}, we first give a graphical explanation of \cref{supeq:caramplitudedecomposed} (see \cref{supfig:paths}).
The \gls{car} coefficient $t_{\mathrm{RL}}^{he}$ is given by the sum of the two lowest order processes $t_1$ and $t_2$ necessary to convert an electron incoming from the left into a hole leaving the heterostructure to the right as also shown in the main text, augmented by the insertion of all possible closed loops, denoted $\Sigma^{\mathrm{L,R}}_1$, $\Sigma^{\up,\dw}_2$ and $\Sigma^{e,h}_3$.

The schematic paths in \cref{supfig:paths} correspond to the expressions
\begin{equation} \label{supeq:lowest_order_car}
\begin{aligned}
 t_{1} &= t^{ee}_{\mathrm{SL}(21)} t^{ee}_{\mathrm{NL}(32)} r^{ee}_{\mathrm{F}(33)} t^{ee}_{\mathrm{NL}(23)} r^{he}_{\mathrm{SL}(22)} t^{hh}_{\mathrm{NL}(32)} t^{hh}_{\mathrm{F}(43)} t^{hh}_{\mathrm{NR}(54)} t^{hh}_{\mathrm{SR}(65)} , \\
 t_{2} &= t^{ee}_{\mathrm{SL}(21)} t^{ee}_{\mathrm{NL}(32)} t^{ee}_{\mathrm{F}(43)} t^{ee}_{\mathrm{NR}(54)} r^{he}_{\mathrm{SR}(55)} t^{hh}_{\mathrm{NR}(45)} r^{hh}_{\mathrm{F}(44)} t^{hh}_{\mathrm{NR}(54)} t^{hh}_{\mathrm{SR}(65)} ,
\end{aligned}
\end{equation}
for the two lowest order paths for CAR, \ie 
\begin{enumerate*}[label=(\roman*)]
 \item transmission through SL, reflection at F, local Andreev reflection at SL under electron-hole conversion, and then transmission all the way through to the right lead ($t_{1}$);
 \item transmission through SL and F, local Andreev reflection at SR under electron-hole conversion, reflection at F followed by tunneling to the right lead ($t_{2}$),
\end{enumerate*}
\begin{equation}
 \begin{aligned}
  \Sigma_{1}^{\mathrm{L}} &= t^{ee}_{\mathrm{NL}(32)} r^{ee}_{\mathrm{F}(33)} t^{ee}_{\mathrm{NL}(23)} r^{he}_{\mathrm{SL}(22)} t^{hh}_{\mathrm{NL}(32)} r^{hh}_{\mathrm{F}(33)} t^{hh}_{\mathrm{NL}(23)} r^{eh}_{\mathrm{SL}(22)} , \\
  \Sigma_{1}^{\mathrm{R}} &= t^{ee}_{\mathrm{NR}(54)} r^{he}_{\mathrm{SR}(55)} t^{hh}_{\mathrm{NR}(45)} r^{hh}_{\mathrm{F}(44)} t^{hh}_{\mathrm{NR}(54)} r^{eh}_{\mathrm{SR}(55)} t^{ee}_{\mathrm{NR}(45)} r^{ee}_{\mathrm{F}(44)} ,
 \end{aligned}
\end{equation}
for the two closed loops in the left (L) and right (R) S-F cavities,
\begin{equation}
 \begin{aligned}
  \Sigma_{2}^{\up} &= t^{ee}_{\mathrm{NL}(32)} t^{ee}_{\mathrm{F}(43)} t^{ee}_{\mathrm{NR}(54)} r^{he}_{\mathrm{SR}(55)} t^{hh}_{\mathrm{NR}(45)} t^{hh}_{\mathrm{F}(34)} t^{hh}_{\mathrm{NL}(23)} r^{eh}_{\mathrm{SL}(22)} , \\
  \Sigma_{2}^{\dw} &= t^{hh}_{\mathrm{NL}(32)} t^{hh}_{\mathrm{F}(43)} t^{hh}_{\mathrm{NR}(54)} r^{eh}_{\mathrm{SR}(55)} t^{ee}_{\mathrm{NR}(45)} t^{ee}_{\mathrm{F}(34)} t^{ee}_{\mathrm{NL}(23)} r^{he}_{\mathrm{SL}(22)} ,
 \end{aligned}
\end{equation}
for the two closed loops between the SL and SR barriers - \ie one involving spin-up electrons and spin-down holes (denoted with superscript $\up$), and the other one built from spin-down electrons and spin-up holes ($\dw$) - and finally
 \begin{equation}
 \begin{aligned}
  \Sigma_{3}^{e} &= t^{ee}_{\mathrm{NL}(32)} t^{ee}_{\mathrm{F}(43)} t^{ee}_{\mathrm{NR}(54)} r^{he}_{\mathrm{SR}(55)} 
				   t^{hh}_{\mathrm{NR}(45)} r^{hh}_{\mathrm{F}(44)} t^{hh}_{\mathrm{NR}(54)} r^{eh}_{\mathrm{SR}(55)} 
				   t^{ee}_{\mathrm{NR}(45)} t^{ee}_{\mathrm{F}(34)} t^{ee}_{\mathrm{NL}(23)} r^{he}_{\mathrm{SL}(22)}
				   t^{hh}_{\mathrm{NL}(32)} r^{hh}_{\mathrm{F}(33)} t^{hh}_{\mathrm{NL}(23)} r^{eh}_{\mathrm{SL}(22)} , \\
  \Sigma_{3}^{h} &= t^{hh}_{\mathrm{NL}(32)} t^{hh}_{\mathrm{F}(43)} t^{hh}_{\mathrm{NR}(54)} r^{eh}_{\mathrm{SR}(55)} 
				   t^{ee}_{\mathrm{NR}(45)} r^{ee}_{\mathrm{F}(44)} t^{ee}_{\mathrm{NR}(54)} r^{he}_{\mathrm{SR}(55)} 
				   t^{hh}_{\mathrm{NR}(45)} t^{hh}_{\mathrm{F}(34)} t^{hh}_{\mathrm{NL}(23)} r^{eh}_{\mathrm{SL}(22)}
				   t^{ee}_{\mathrm{NL}(32)} r^{ee}_{\mathrm{F}(33)} t^{ee}_{\mathrm{NL}(23)} r^{he}_{\mathrm{SL}(22)} ,
 \end{aligned}
\end{equation}
for the loops between the S regions with an additional detour in each of the cavities (see \cref{supfig:paths}). The superscript refers to the particle type along the long paths connecting the S regions.

\vspace{\baselineskip}

In order to derive Eq.~(3) of the main text, one can now analyze the energy dependence of $\bm{\varSigma}$ by making use of particle hole symmetry. The terms $\Sigma_{1}^{\mathrm{L}}$ and $\Sigma_{1}^{\mathrm{R}}$ involve all four modes and all possible reflections and thus are invariant under charge conjugation, \ie
\begin{equation} \label{supeq:phsSigmaLR}
 \Sigma_{1}^{\mathrm{L/R}}(E) = \left( \Sigma_{1}^{\mathrm{L/R}}(-E) \right)^{*} ,
\end{equation} 
which implies that the real and imaginary parts fulfill 
\begin{align} 
 \real \left[ \Sigma_{1}^{\mathrm{L/R}}(E) \right] = \real \left[ \Sigma_{1}^{\mathrm{L/R}}(-E) \right], &&
 \imag \left[ \Sigma_{1}^{\mathrm{L/R}}(E) \right] = - \imag \left[ \Sigma_{1}^{\mathrm{L/R}}(-E) \right] .
\end{align} 
By contrast, $\Sigma_{2}^{\up/\dw}$ and $\Sigma_{3}^{e/h}$ are charge conjugated partners of one another, such that
\begin{align} \label{supeq:phsSigmaupdowneh}
 \Sigma_{2}^{\up}(E) = \left( \Sigma_{2}^{\dw}(-E) \right)^{*} ,
 &&  
 \Sigma_{3}^{e}(E) = \left( \Sigma_{3}^{h}(-E) \right)^{*} ,
\end{align} 
and hence
\begin{equation}
\begin{alignedat}{3}
 \real \left[ \Sigma_{2}^{\up}(E) \right] &= \real \left[ \Sigma_{2}^{\dw}(-E) \right]   ,
 \qquad & \qquad
 \imag \left[ \Sigma_{2}^{\up}(E) \right] &= - \imag \left[ \Sigma_{2}^{\dw}(-E) \right] ,
 \\
 \real \left[ \Sigma_{3}^{e\vphantom{\up}}(E) \right] &= \real \left[ \Sigma_{3}^{h\vphantom{\up}}(-E) \right] ,
 \qquad & \qquad
 \imag \left[ \Sigma_{3}^{e\vphantom{\up}}(E) \right] &= - \imag \left[ \Sigma_{3}^{h\vphantom{\up}}(-E) \right] .
\end{alignedat} 
\end{equation}

Next, we consider the complex valued functions $u(E), w(E), z(E)$ defined as
\begin{equation}
\begin{aligned}
u(E) &= \Sigma_{1}^{\mathrm{L}}(E) + \Sigma_{1}^{\mathrm{R}}(E) - \Sigma_{1}^{\mathrm{L}}(E) \Sigma_{1}^{\mathrm{R}}(E) ,
\\
w(E) &= \Sigma_{2}^{\mathrm{\up}}(E) + \Sigma_{2}^{\dw}(E) - \Sigma_{2}^{\up}(E) \Sigma_{2}^{\dw}(E) ,
\\ 
z(E) &= \Sigma_{3}^{\mathrm{e}}(E) + \Sigma_{3}^{h}(E) ,
\end{aligned} 
\end{equation}
such that $u+w+z=\bm{\varSigma}$. Importantly,
\begin{align} \label{supeq:phsvarSigmaparts}
 u^{*}(E) = u(-E) , \quad w^{*}(E) = w(-E), \quad z^{*}(E) = z(-E),
\end{align} 
because of \cref{supeq:phsSigmaLR,supeq:phsSigmaupdowneh}. \cref{supeq:phsvarSigmaparts} immediately leads to the crucial result
\begin{equation}
\left[ \bm{\varSigma}(E) \right]^{*} = \bm{\varSigma}(-E) ,
\end{equation} 
demonstrating that $\abs{\bm{\varSigma}}$ is indeed even in $E$.
Going back to the full crossed Andreev reflection coefficient, we can rewrite \cref{supeq:caramplitudedecomposed} as
\begin{equation} \label{supeq:caramplitude_realdenom}
\begin{aligned}
 t_{\mathrm{RL}}^{he} &= \left( t_1 + t_2 \right) \, \frac{1 - \bm{\varSigma}^{*}}{1 + \abs{\bm{\varSigma}}^{2} - \bm{\varSigma} - \bm{\varSigma}^{*} } ,
\end{aligned}
\end{equation} 
where the denominator $d(E)=1 + \abs{\bm{\varSigma}(E)}^{2} - \bm{\varSigma}(E) - \left( \bm{\varSigma}(E) \right)^{*}$ is real and obeys
\begin{equation} \label{supeq:caramplitude_realdenom_denom}
 d^{*}(E) = d(-E) \qquad \Rightarrow \quad \abs{d(E)}^{2} = \abs{d(-E)}^{2} ,
\end{equation} 
while finally the numerator of the second term in \cref{supeq:caramplitude_realdenom} is $n(E) = 1 - \left( \bm{\varSigma}(E) \right)^{*}$ with the property
\begin{equation} \label{supeq:caramplitude_realdenom_num}
 n^{*}(E) = n(-E) \qquad \Rightarrow \quad \abs{n(E)}^{2} = \abs{n(-E)}^{2} .
\end{equation}
In conclusion, the modulus of the contribution of all higher order corrections given by $1/(1 - \bm{\varSigma}) = d/n$ is indeed even in energy.

\vspace{\baselineskip}

As the final step, we turn to the first term in \cref{supeq:caramplitude_realdenom} responsible for the interference effect.
From \cref{supeq:lowest_order_car}, we first write 
\begin{gather} \label{supeq:lowest_order_car_factored}
    t_1 + t_2 = t^{ee}_{\mathrm{SL}(21)} t^{ee}_{\mathrm{NL}(32)}   t^{hh}_{\mathrm{NR}(54)} t^{hh}_{\mathrm{SR}(65)}  
    \left(   r^{ee}_{\mathrm{F}(33)} t^{ee}_{\mathrm{NL}(23)} r^{he}_{\mathrm{SL}(22)} t^{hh}_{\mathrm{NL}(32)} t^{hh}_{\mathrm{F}(43)}
    +  t^{ee}_{\mathrm{F}(43)} t^{ee}_{\mathrm{NR}(54)} r^{he}_{\mathrm{SR}(55)} t^{hh}_{\mathrm{NR}(45)} r^{hh}_{\mathrm{F}(44)} \right) .
\end{gather}
Now, using the explicit form of the coefficients we provide in \cref{supeq:coefficients} below, we have 
\begin{equation} \label{supeq:relations_coefficients}
\begin{gathered}
    t^{hh}_{\mathrm{F}(43)} = t^{ee}_{\mathrm{F}(43)}, \qquad r^{ee}_{\mathrm{F}(33)} = r^{hh}_{\mathrm{F}(44)}, \qquad r^{he}_{\mathrm{SR}(55)} = \ee^{\ii \phi} \, r^{he}_{\mathrm{SL}(22)} \\
    t^{ee}_{\mathrm{NL}(23)} = t^{hh}_{\mathrm{NL}(32)} = \ee^{\ii d_\mathrm{NL} E}, \qquad \text{and} \qquad t^{ee}_{\mathrm{NR}(54)} = t^{hh}_{\mathrm{NR}(45)} = \ee^{\ii d_\mathrm{NR} E}.
\end{gathered}
\end{equation}
Thus, \cref{supeq:lowest_order_car_factored} simplifies to 
\begin{gather} 
    t_1 + t_2 = t^{ee}_{\mathrm{SL}(21)} t^{ee}_{\mathrm{NL}(32)}   t^{hh}_{\mathrm{NR}(54)} t^{hh}_{\mathrm{SR}(65)}  r^{ee}_{\mathrm{F}(33)} r^{he}_{\mathrm{SL}(22)} t^{ee}_{\mathrm{NR}(54)}
    \left(  \ee^{2 \ii d_\mathrm{NL} E }
    +  \ee^{2 \ii d_\mathrm{NR} E + \ii \phi}  \right) ,
\end{gather}
which is of the form
\begin{equation} \label{supeq:lowest_order_car_final}
    t_1 + t_2 = 2 \abs{t(E)} \ee^{\ii \tau(E)} \, \cos\left(\Delta\varphi \right/2) .
\end{equation}
Here, we have introduced the phase difference between the two paths $\Delta\varphi = \phi + 2 (d_{\mathrm{NR}} - d_{\mathrm{NL}}) E$, an unimportant global phase $\tau$, and the modulus of a product of scattering coefficients fulfilling $\abs{t(E)} = \abs{t(-E)}$ (see below).
Combining \cref{supeq:caramplitude_realdenom,supeq:caramplitude_realdenom_denom,supeq:caramplitude_realdenom_num,supeq:lowest_order_car_final}, we thus show that the absolute squared of the CAR coefficient can be written in the form
\begin{equation}
 \abs{t^{he}_{\mathrm{RL}}}^2 = T^{he}_{\mathrm{RL}}(E) = \gamma(E,\phi) \cos^{2} \left[ \phi/2 + \left( d_{\mathrm{NR}} - d_{\mathrm{NL}}\right) E \right] ,
\end{equation} 
where $\gamma(E,\phi)$ is even in $E$ and given by (we have restored the dependence on the phase difference $\phi$)
\begin{equation}
 \gamma(E,\phi) = \frac{4 \abs{t(E,\phi)}^{2} \abs{ n(E,\phi])}^{2} }{ \abs{d(E,\phi)}^{2} } ,
\end{equation} 
as stated in Eq.~(3) in the main text. 

\vspace{\baselineskip}

In order to show that $\abs{t(E)}$ is indeed even in $E$, we define $A_\mathrm{S}(E) = \arccosh (E/\Delta)$, $A_\mathrm{F}(E) = \arccosh (E/m_0)$ and 
\begin{align*}
    \Omega_\mathrm{S}(E) = 
    \begin{cases}
    \sqrt{E^2 - \Delta^2}       &\text{for } E > \Delta \\
    \ii \sqrt{\Delta^2 - E^2}   &\text{for } -\Delta < E < \Delta \\
    - \sqrt{E^2 - \Delta^2}       &\text{for } E < -\Delta 
    \end{cases} ,
    \quad 
     \Omega_\mathrm{F}(E) = 
    \begin{cases}
    \sqrt{E^2 - m_0^2}       &\text{for } E > m_0 \\
    \ii \sqrt{m_0^2 - E^2}   &\text{for } -m_0 < E < m_0 \\
    - \sqrt{E^2 - m_0^2}       &\text{for } E < -m_0 
    \end{cases}.
\end{align*}
The coefficients are then found to be [together with \cref{supeq:relations_coefficients}]
\begin{equation} \label{supeq:coefficients}
\begin{alignedat}{3} 
    t^{ee}_{\mathrm{SL}(21)}(E) &= \frac{\sinh\left[ A_\mathrm{S}(E) \right]}{\sinh \left[ A_\mathrm{S}(E) - \ii d_\mathrm{SL} \Omega_\mathrm{S}(E) \right] }  ,  \qquad &
    r^{he}_{\mathrm{SL}(22)}(E) &= -\ii \frac{\sin\left[  d_{\mathrm{SL}}  \Omega_\mathrm{S}(E) \right]}{\sinh \left[ A_\mathrm{S}(E) - \ii d_\mathrm{SL} \Omega_\mathrm{S}(E) \right] }  
    , \\
    t^{hh}_{\mathrm{SR}(65)}(E) &= \frac{\sinh\left[ A_\mathrm{S}(E) \right]}{\sinh \left[ A_\mathrm{S}(E) - \ii d_\mathrm{SR} \Omega_\mathrm{S}(E) \right] } , &
    r^{ee}_{\mathrm{F}(33)}(E) &= -\ii \frac{\sin\left[  d_{\mathrm{F}}  \Omega_\mathrm{F}(E) \right]}{\sinh \left[ A_\mathrm{F}(E) - \ii d_\mathrm{F} \Omega_\mathrm{F}(E) \right] }   
    \\
   t^{ee}_{\mathrm{F}(43)}(E) &= \frac{\sinh\left[ A_\mathrm{F}(E) \right]}{\sinh \left[ A_\mathrm{F}(E) - \ii d_\mathrm{F} \Omega_\mathrm{F}(E) \right] } , &  t^{ee}_{\mathrm{NL}(32)}(E) &=  \ee^{\ii d_\mathrm{NL} E}, \quad   t^{hh}_{\mathrm{NR}(54)}(E) =  \ee^{\ii d_\mathrm{NR} E}.
\end{alignedat}
\end{equation}
From \cref{supeq:coefficients} one can check that the modulus of all coefficients is an even function of $E$, and thus $\abs{t(E)}$ must be even as well.

\section{Green function techniques and superconducting pairing}

The retarded Green function is a $4\times4$-matrix in Nambu and spin space related to the Hamiltonian in Eq.~(1) of the main text. In the basis $\Psi(x) = (\psi_\up, \psi_\dw, \psi^\dagger_\dw, - \psi^\dagger_\up)$, it takes the form
\begin{equation*}
G^R(x,x',E) = 
\begin{pmatrix}
G_{ee}^R    &&  G_{eh}^R \\[2pt]
G_{he}^R    &&  G_{hh}^R
\end{pmatrix} ,
\qquad
G^R_{eh}(x,x',E) = 
\begin{pmatrix}
[G^R_{eh}]^{\up\dw}    &&  [G^R_{eh}]^{\up\up} \\[2pt]
[G^R_{eh}]^{\dw\dw}    &&  [G^R_{eh}]^{\dw\up}
\end{pmatrix}
\qquad
G^R_{he}(x,x',E) = 
\begin{pmatrix}
[G^R_{he}]^{\dw\up}    &&  [G^R_{he}]^{\dw\dw} \\[2pt]
[G^R_{he}]^{\up\up}    &&  [G^R_{he}]^{\up\dw}
\end{pmatrix}
.
\end{equation*}
Note that due to the choice of basis the functions $[G^{R}_{eh}]^{\up\up/\dw\up}$ and $[G^{R}_{he}]^{\up\up/\up\dw}$ are the negative of the corresponding correlator. 
Following Ref.~\onlinecite{Keidel_2018}, we can use particle-hole symmetry to relate the electron-hole and hole-electron blocks. Explicitly, particle-hole symmetry relates the blocks according to
\begin{gather*}
\mathbf{P} \, G^{R}(x,x',E) \, \mathbf{P}^{-1} = - G^{R}(x,x', -E) ,
\end{gather*}
with $\mathbf{P} = \ty \sy \mathbf{K}$ the charge conjugation operator and $\mathbf{K}$ complex conjugation. This translates into
\begin{equation*} \label{eq:relation_Geh_Ghe}
G^R_{eh}(x,x',-\omega) = \sigma_2 \left( G^R_{he}(x,x',\omega \right)^* \sigma_2 ,
\end{equation*}
which, anticipating what we need in the following, can be recast as
\begin{equation} \label{eq:relation_Geh_Ghe_explicit}
G^R_{he}(x,x',E) = 
\begin{pmatrix}
[G^R_{he}]^{\dw\up}    &&  [G^R_{he}]^{\dw\dw} \\[2pt]
[G^R_{he}]^{\up\up}    &&  [G^R_{he}]^{\up\dw}
\end{pmatrix}
=  \sigma_2 \left( G^R_{eh}(x,x',-E) \right)^* \sigma_2
=
\begin{pmatrix}
\left(  [G^R_{eh}]^{\dw\up} \right)^*   &&  - \left( [G^R_{eh}]^{\dw\dw} \right)^* \\[2pt]
- \left(  [G^R_{eh}]^{\up\up} \right)^*   &&  \left( [G^R_{eh}]^{\up\dw} \right)^*
\end{pmatrix}(x,x',-E) .
\end{equation}

The spin components of the anomalous part of the Green function are extracted by writing
\begin{equation}
G^R_{eh}(x,x',E) = \sum_{k=0}^{3} f_k(x,x',E) \sigma_k
\end{equation}
where $\sigma_k$ are the Pauli matrices in spin space. The singlet is given by the $k=0$ component. Inverting results in 
\begin{align}
f_0 &= \frac{1}{2} \left( [G^R_{eh}]^{\up\dw}  +  [G^R_{eh}]^{\dw\up} \right)   \\
f_1 &= \frac{1}{2} \left( [G^R_{eh}]^{\up\up}  +  [G^R_{eh}]^{\dw\dw} \right)  \\
f_2 &= \frac{1}{2} \ii \left( [G^R_{eh}]^{\up\up}  -  [G^R_{eh}]^{\dw\dw} \right) \\ 
f_3 &= \frac{1}{2} \left( [G^R_{eh}]^{\up\dw}  -  [G^R_{eh}]^{\dw\up} \right) .
\end{align}
Note that our choice of basis leads to positive sign for $f_0$ and the relative minus sign for $f_3$. We also define the shorthand notation $f_+ = [G^R_{eh}]^{\up\up} = f_1 - \ii f_2$ and $f_- = [G^R_{eh}]^{\dw\dw} = f_1 + \ii f_2$ for the equal-spin pairings.

%
%


\begin{figure}[h]
	\begin{center}
		\includegraphics[width=0.99\textwidth]{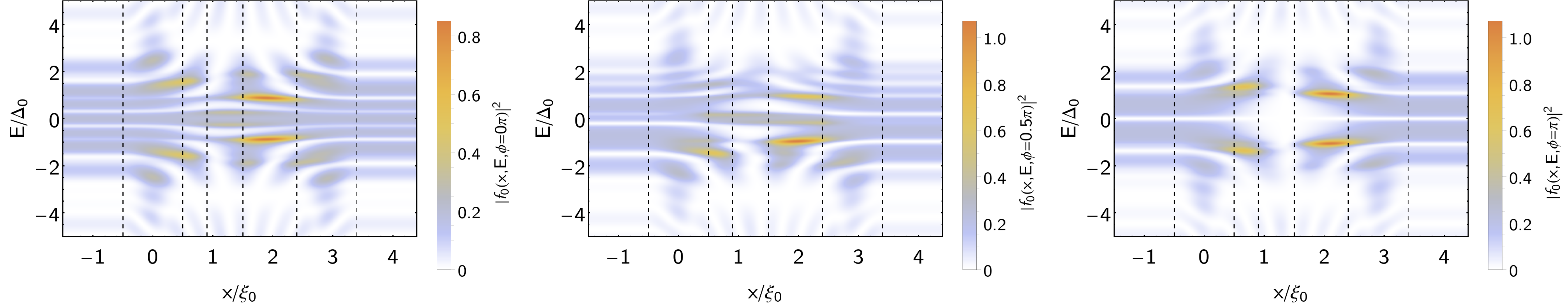} \\
		\includegraphics[width=0.99\textwidth]{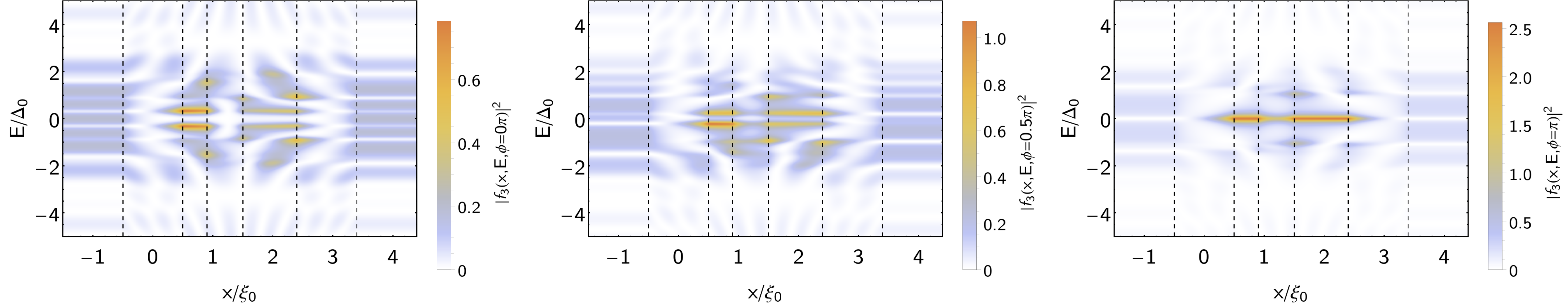} \\
		\includegraphics[width=0.99\textwidth]{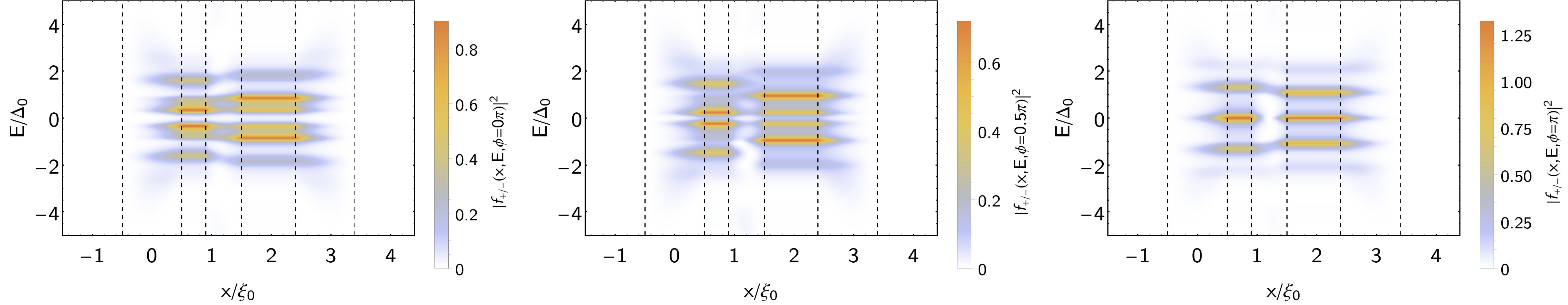}
		\caption{\label{fig:localpairings}
			Plot of the spin components of the local anomalous Green function $G_{eh}$ as a function of energy and position, from top to bottom: singlet, $t_0$-triplet, $\up\up/\dw\dw$-triplet (the latter two are equal).
		}
	\end{center}
\end{figure}


We can now analyze the equal-spin components of the Green function $[G^R_{eh}]^{\up\up/\dw\dw}$ and $f_0$ and $f_3$ (see \cref{fig:localpairings}), or simply all components $f_k$. 

Another option is to consider the polarization vector
\begin{equation} \label{eq:definitionpvec}
\Vec{p} = \ii \Vec{f} \times \Vec{f}^* = 
\ii
\begin{pmatrix}
f_2 f_3^* - f_3 f_2^* \\
- f_1 f_3^* + f_3 f_1^* \\
f_1 f_2^* - f_2 f_1^*
\end{pmatrix} 
= - 2
\begin{pmatrix}
\imag{f_2 f_3^*} \\
\imag{f_3 f_1^*} \\
\imag{f_1 f_2^*}
\end{pmatrix}    ,
\end{equation}
where $\Vec{f} = (f_1,f_2,f_3)$~\cite{Sigrist_RMP,Legget_RMP}. 
Similarly, we can decompose the hole-electron block of the anomalous Green function through 
\begin{equation}
G^R_{he}(x,x',E) = \sum_{k=0}^{3} f'_k(x,x',E) \sigma_k  ,
\end{equation}
which implies
\begin{align}
f'_0 &= \frac{1}{2} \left( [G^R_{he}]^{\dw\up}  +  [G^R_{he}]^{\up\dw} \right)   \\
f'_1 &= \frac{1}{2} \left( [G^R_{he}]^{\dw\dw} + [G^R_{he}]^{\up\up}  \right)  \\
f'_2 &= \frac{1}{2} \ii \left( [G^R_{he}]^{\dw\dw} - [G^R_{he}]^{\up\up}  \right) \\ 
f'_3 &= \frac{1}{2} \left( [G^R_{he}]^{\dw\up} - [G^R_{he}]^{\up\dw}  \right) .
\end{align}


\begin{figure}[h]
	\begin{center}
		\includegraphics[width=0.99\textwidth]{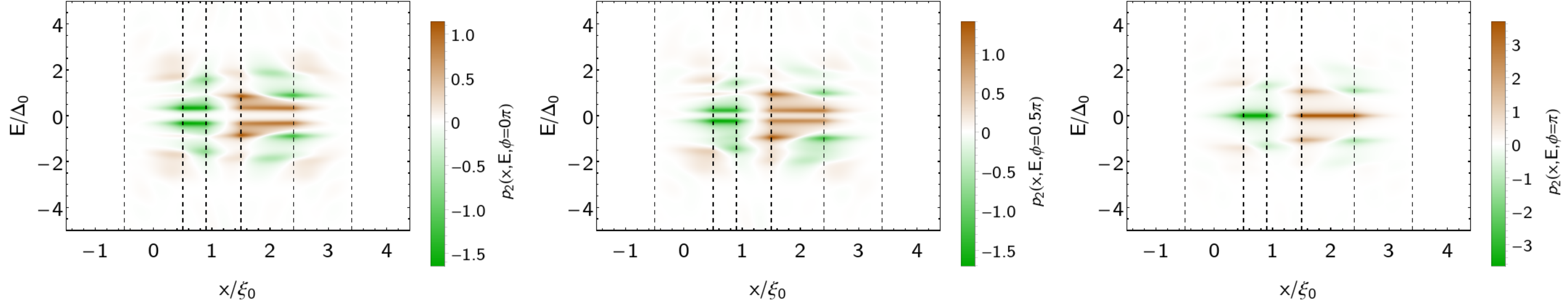} \\
		\includegraphics[width=0.99\textwidth]{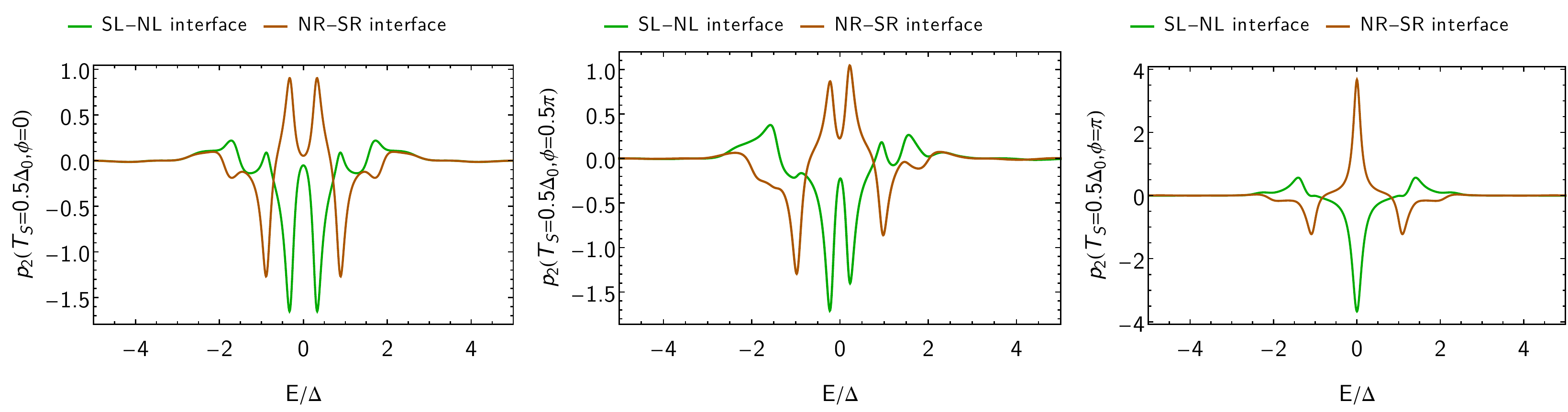} 
		\caption{\label{fig:localpvec}
			Plot of the $p_2$-component of the polarization vector $\Vec{p}$ from local pairing for phase differences $0$, $\pi/2$, and $\pi$ from left to right.
			Top row: map of the modulus as a function of energy and position.
			Bottom row: cuts at the SL-NL and NR-SR interfaces. Note the symmetry in energy for phase differences $0, \pi$. 
		}
	\end{center}
\end{figure}


Put together with the relation between the blocks in \cref{eq:relation_Geh_Ghe_explicit}, the vectors $\Vec{f}$ and $\Vec{f'}$ are related via
\begin{equation} \label{eq:relation_ffprime}
\Vec{f'} = 
\begin{pmatrix}
f'_1 \\ f'_2 \\ f'_3
\end{pmatrix}
=  - 
\begin{pmatrix}
f_1^*(-E) \\ f_2^*(-E) \\f_3^*(-E)
\end{pmatrix} ,
\end{equation}
such that the polarization vector for the hole-electron block fulfills
\begin{equation}
\Vec{p'} = \ii \Vec{f'} \times \Vec{f'}^* =-2  \begin{pmatrix}
\imag \left[ f'_2 (f'_3)^* \right] \\
\imag \left[ f'_3 (f'_1)^* \right] \\
\imag \left[ f'_1 (f'_2)^* \right]
\end{pmatrix} 
= -2 
\begin{pmatrix}
\imag{f_2^* f_3}  \\
\imag{f_3^* f_1}  \\
\imag{f_1^* f_2} 
\end{pmatrix} (-E)
= - \Vec{p}(-E).
\end{equation}

The local equal-spin pairings are exactly equal, i.e. $[G^R_{eh}]^{\up\up}  =  [G^R_{eh}]^{\dw\dw}$, meaning that $\mean{\psi^\dagger_{\up}\psi^\dagger_{\up}}=-\mean{\psi^\dagger_{\dw}\psi^\dagger_{\dw}}$, which directly implies that $f_2 = 0$. From \cref{eq:definitionpvec} we immediately see that $p_1 = p_3 = 0$, i.e. $\Vec{p}$ points in the $y$-direction. 
We show in \cref{fig:localpairings} the squared moduli of $f_{\pm}, f_3, f_0$ for phase differences $0, 0.5\pi, \pi$. 
The resulting polarization vector $\Vec{p}$, more precisely the only nonzero component $p_2$, is shown in \cref{fig:localpvec}.

\section{Temperature dependence of the noise in the right lead} 

\begin{figure*}[ht]
	\includegraphics[scale=0.99]{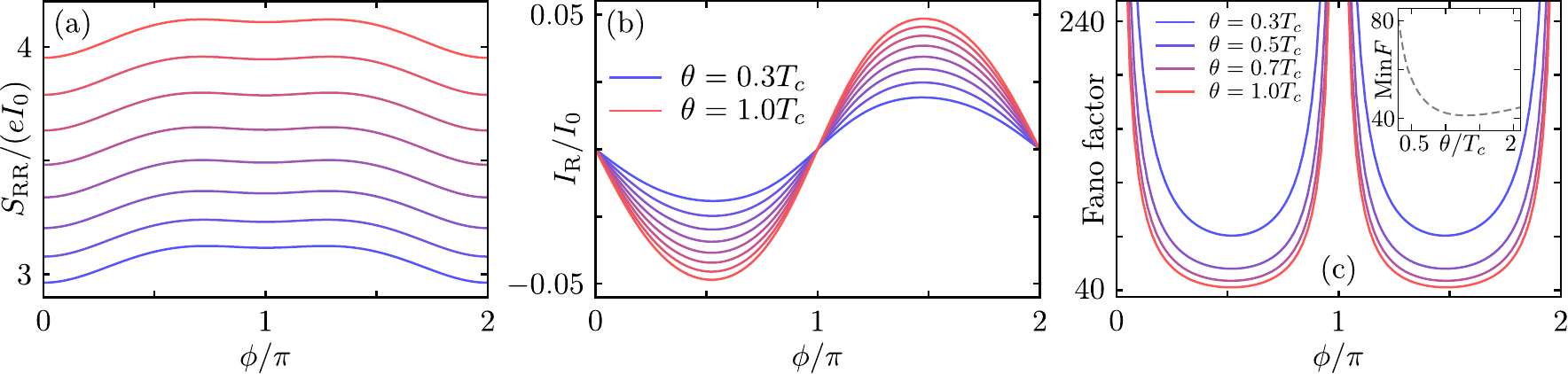}
	\caption{\label{supfig:noiseproperties}
		(a) Noise and (b) average current in the right lead as a function of the phase difference $\phi$ and for $\theta$ between $0.3 T_{c}$ and $1.0 T_{c}$ in equidistant steps of $0.1 T_c$. (c) Fano factor for selected temperature differences. The inset shows the minimum of $F$ as a function of $\theta$. 
		Unless specified otherwise, we use the parameters $d_{\mathrm{SL}} \!=\! 
		d_{\mathrm{SR}} \!=\! \xi_0$, $d_{\mathrm{FM}} \!=\! 0.6 \xi_0$, $d_{\mathrm{NL}} \!=\! 0.4 \xi_0$, $d_{\mathrm{NR}} \!=\! 
		0.9 \xi_0$, $m_{0} \!=\! 1.5 \Delta_{0}$, $T_{0} \!=\! 0.5 T_c $, $\phi\!=\!\pi/2$, and $T_c \!=\! \Delta_{0}$.
	}
\end{figure*}

As we describe in the main text, the noise in the right lead, $S_{\mathrm{RR}}$, defined in Eq. (5) of the main text, is almost independent of the phase difference, see \cref{supfig:noiseproperties}(a). This is a result of the noise being dominated by the thermal contribution, which is caused by normal processes that do not experience interference. 
On the other hand, the thermoelectric current in the right lead, $I_{\mathrm{R}}$, depends on the phase difference $\phi$ between the superconductors, due to the interference effect of \gls{car} processes, see \cref{supfig:noiseproperties}(b). 
Consequently, the Fano factor strongly depends on the phase difference, see \cref{supfig:noiseproperties}(c), and is minimum when the current is maximum. 
Moreover, as the inset of \cref{supfig:noiseproperties}(c) indicates, the minimum of the Fano factor is rather stable for temperature biases $\theta\sim T_c$. 


In the main text, we focus on the behavior of the current fluctuations and the Fano factor as a function of the temperature difference $\theta$ at fixed base temperature $T_0$. Here, we provide some information about the dependence on $T_0$. 
We fix the phase difference between the superconductors to be at the optimal operating value $\phi = \pi/2$. The temperature difference is set to be $\theta = 2 T_0$ (plotted in cyan) and $\theta = T_c$ (orange). Note that at $T_0 = T_c/2$, both cases coincide and we also recover the results of the main text.


\begin{figure}[t]
	\begin{center}
		\includegraphics[scale=0.99]{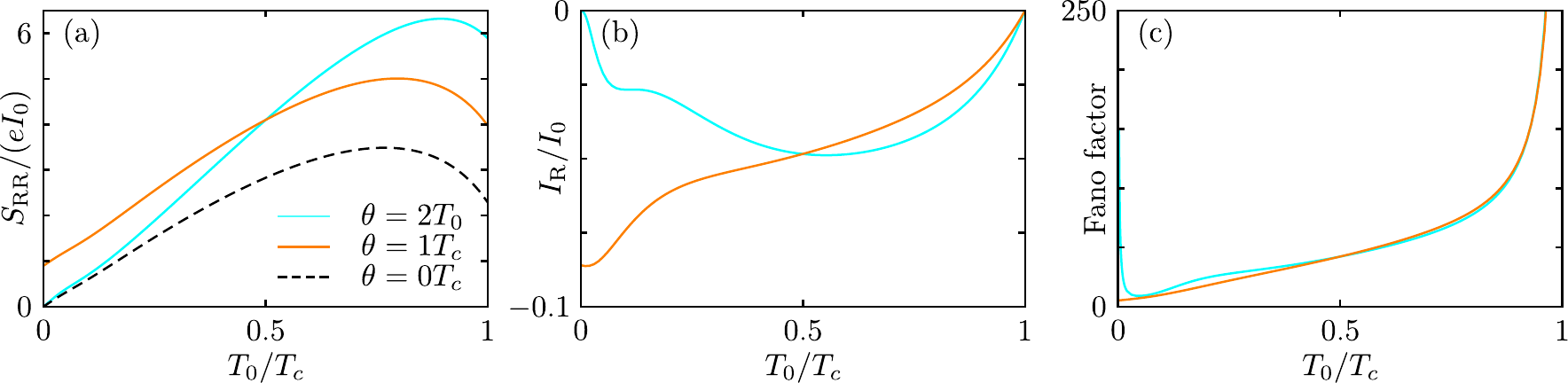}
	\end{center}
	\caption{\label{supfig:noise_T0} 
		(a) Noise, (b) average current, and (c) Fano factor in the right lead as a function of the base temperature $T_0$ with temperature differences $\theta = 2 T_0$ (cyan) and $\theta = T_c$ (orange) fixed. The dashed black line in (a) shows the noise in the equilibrium case for $\theta = 0$. 
		The phase difference is $\phi=\pi/2$ for all panels, and the other parameters are chosen as in the main text.
	}
\end{figure}


\cref{supfig:noise_T0} (a) shows the noise in the right lead $S_{\mathrm{RR}}$. For all choices for the temperature gradient $\theta$ the fluctuations do not grow monotonically up to $T_0 = T_c$, but reach a maximum below that value. 
This is connected to the fact that the scattering problem and thus the conductance change dramatically when the two superconducting barriers are removed. 
Note that there are several contributions to the noise in setups like this, namely an equilibrium contribution due to finite temperatures, a non-equilibrium contribution due to a temperature gradient and, additionally, a contribution due to the thermoelectrically created nonlocal charge current. The interplay of these sources of fluctuations has been carefully studied in~\cite{Lumbroso2018}.
In \cref{supfig:noise_T0} (b), we plot the total current in the right lead produced by the temperature gradient. For smaller values of $T_0$, the current for $\theta = T_c$ is of course much larger than for $\theta = 2 T_0$, but for both cases it is not dominated by Andreev contributions below $T_0 \approx T_c/2$. The quite sharp drop of the $\theta = T_c$ case for $T_0$ smaller than $T_c/2$ is mostly due to the vanishing of the normal contribution.
Finally, \cref{supfig:noise_T0} (c) displays the Fano factor $F = S_{\mathrm{RR}}/\abs{2e I_{\mathrm{R}}}$ resulting from noise and average current in \cref{supfig:noise_T0} (a) and (b). 
For small base temperatures, Fano factors below 10 are possible. However, note that a dominating Andreev contribution to the current requires $T_0$ to be of the order of $T_c/2$, which increases the Fano factor at the optimal working point of our device.

\end{document}